\documentclass{article}
\usepackage[a4paper, total={6in, 8in}]{geometry}

\usepackage{graphicx} 
\usepackage{amsmath}
\usepackage{amssymb}

\usepackage{ulem}
\usepackage{xcolor}

\newtheorem{theorem}{Theorem}

\title{Lanczos Method for QRPA Strength Functions in Atomic Nuclei}

\author{%
Dong Min Roh\thanks{Computational Research Division, Lawrence Berkeley National Laboratory, Berkeley, California 94720, United States}
\and
Matthew L. Dai\thanks{Department of Physics and Astronomy, University of North Carolina, Chapel Hill, NC 27599-3255, United States}
\and
Jonathan Engel\footnotemark[2]
\and
Chao Yang\footnotemark[1]
}
\date{June 2026}

\begin{document}

\maketitle

\begin{abstract}
We present a symmetric Lanczos method for computing charge-changing QRPA strength functions in atomic nuclei. Starting from the finite-amplitude-method formulation of the QRPA linear-response problem, we derive equivalent spectral representations and, in the real case, a reduced eigenvalue problem involving the matrix products $MK$ and $KM$, where $M\equiv A+B$ and $K\equiv A-B$ are formed from the usual QRPA matrices $A$ and $B$. The resulting formulation enables a matrix-free Lanczos approximation of the Lorentzian-smeared strength function over a broad energy interval from a single Krylov run, in contrast to conventional frequency-by-frequency response calculations. Numerical tests for $^{112}$Sn and $^{150}$Nd first show that GMRES reproduces the converged iterative FAM strength profiles while requiring fewer iterations. Using GMRES as the frequency-by-frequency reference, we then show that the Lanczos approximation reproduces the same strength profiles with reduced overall cost. These results indicate that symmetric Lanczos projection provides an efficient and accurate approach for QRPA strength-function calculations when spectral information is required over an extended frequency range.
\end{abstract}

\section{Introduction}

Charge-changing nuclear response functions are important in nuclear-structure and weak-interaction phenomenology.
They 
enter the description of $\beta$ decay and double-$\beta$ decay, and provide information about collective excitations and weak rates in nuclei across the chart of nuclides \cite{ring1983nuclear,engel2017review}.
For medium-mass and heavy open-shell nuclei, especially in deformed systems, the quasiparticle random-phase approximation (QRPA) built on nuclear energy-density functionals provides a practical microscopic framework for describing such excitations and transition strengths \cite{ring1983nuclear,nakatsukasa2016tddft}.

In principle, QRPA strength functions can be obtained from the explicit QRPA matrix or from a full eigendecomposition of the linear-response problem.
In practice, however, the matrix dimension grows rapidly with the size of the quasiparticle space, and explicit construction becomes prohibitively expensive for heavy and deformed nuclei.
The finite amplitude method (FAM) addresses this difficulty by computing the QRPA response through induced fields, without assembling the full QRPA matrix explicitly \cite{nakatsukasa2007fam,avogadro2011finite}.
The FAM has become an effective tool for self-consistent linear-response calculations, including applications to superfluid and deformed nuclei and, in particular, to charge-changing transitions in axially deformed systems \cite{hinohara2013low,mustonen2014finite,ney2020global}.

Despite their usefulness, conventional FAM-based calculations still proceed one frequency at a time. For each complex frequency on the chosen contour, one must solve a new linear system.
This feature is well suited to pointwise evaluation of the response, but it can become costly when the goal is to reconstruct the strength function over a broad energy interval.
That observation motivates Krylov-based approaches that capture spectral information more globally.
These have appeared in iterative Arnoldi methods for strength functions, in spectral-density approximation techniques, and in recent polynomial and kernel-based approaches to QRPA response problems \cite{toivanen2010linear,lin2016approximating,brabec2015efficient,bjelvcic2022chebyshev,bjelvcic2025computing}.

In this paper, we develop a symmetric Lanczos framework for approximating charge-changing QRPA strength functions.
Starting from the FAM linear-response equation, we derive equivalent spectral formulations and a reduced eigenvalue problem involving the matrices $MK$ and $KM$.
These reformulations make it possible to apply a symmetric Lanczos process with respect to the appropriate inner product and to approximate the smeared strength function over a wide frequency interval from a single Krylov run, rather than from separate solves at each frequency.
Our goal is to retain the main efficiency advantages of FAM-style matrix-free calculations while exploiting the broader spectral information available through Lanczos projection.

The paper is organized as follows.
Section~\ref{sec:qrpa_formulation} develops the QRPA formulations of the strength function, including the eigendecomposition-based and reduced-eigenproblem representations.
Section~\ref{sec:computational_methods} presents the frequency-by-frequency solvers and the symmetric Lanczos approximation.
Section~\ref{sec:numerical_results} reports numerical results for $^{112}$Sn and $^{150}$Nd, together with runtime comparisons.

\paragraph{Notation.}
Throughout the paper, matrices and vectors may be real or complex, depending on the formulation under consideration.
For a matrix $A\in\mathbb{C}^{m\times n}$, we write $A^\dag$ for the Hermitian conjugate, $A^T$ for the transpose, and $A^*$ for the element-wise complex conjugate.
The identity matrix is denoted by $I$, with its size understood from context.
When block matrices are used, their dimensions are likewise determined by the surrounding equations.
We reserve $\omega$ for the physical excitation energy and write $\omega_\gamma=\omega+i\gamma$ when a finite Lorentzian smearing width $\gamma>0$ is introduced.
Unless stated otherwise, vectors are written as columns, and the Euclidean inner product is understood by default; when the analysis requires a different metric, such as the $K$-inner product used in the Lanczos formulation, we say so explicitly.

\section{QRPA Formulation of the Strength Function}\label{sec:qrpa_formulation}

This section summarizes the QRPA equations and spectral reformulations that underlie the strength-function calculations developed later. Although the derivation passes through the linear response equation and the response function, the quantity of interest throughout is the charge-changing QRPA strength function. The iterative and Lanczos-based algorithms used to evaluate it are deferred to Section~\ref{sec:computational_methods}.

\subsection{The FAM Equations and the Strength Function}

The nuclear linear response, a function of frequency $\omega$, is genrated by a weak external time-dependent field $\hat{F}(t)$ of the form 
\begin{equation}\label{eq:Ft}
    \hat{F}(t) = \eta \left(\hat{F}(\omega)e^{-i\omega t} + \hat{F}^\dag(\omega)e^{i\omega t}\right) \,,
\end{equation}
where $\eta$ is a small real parameter.
Neglecting two-quasiparticle operators that do not contribute to the QRPA, the field can be written in the quasiparticle basis as
\begin{equation}\label{eq:Fw}
    \hat{F}(\omega) = \frac{1}{2}\sum_{\mu\nu}\left(F_{\mu\nu}^{20}(\omega)\hat{\alpha}_\mu^\dag\hat{\alpha}_\nu^\dag+F_{\mu\nu}^{02}(\omega)\hat{\alpha}_\mu\hat{\alpha}_\nu\right),
\end{equation}
where $F_{\mu\nu}^{20}(\omega)$ and $F_{\mu\nu}^{02}(\omega)$ are defined, e.g., in Ref.\ \cite{ring1983nuclear}, and $\hat{\alpha}_\mu^\dag$ and $\hat{\alpha}_\mu$ are quasiparticle creation and annihilation operators, respectively.
For charge-changing applications of the finite-amplitude method in deformed nuclei, we refer in particular to \cite{avogadro2011finite,mustonen2014finite,hinohara2013low}.

In the QRPA, the time evolution of the quasiparticle operators under the external field $\hat{F}(t)$ is governed by the time-dependent Hartree--Fock--Bogoliubov (TDHFB) equation:
\begin{equation}\label{eq:TDHFB_eq}
    i\partial_t\hat{\alpha}_\mu(t) = \left[\hat{H}(t)+\hat{F}(t), \hat{\alpha}_\mu(t)\right],
\end{equation}
where the quasiparticle operators acquire small-amplitude oscillations of the form
\begin{align}
    \hat{\alpha}_\mu(t) &= \left(\hat{\alpha}_\mu+\delta\hat{\alpha}_\mu(t)\right)e^{iE_\mu t}, \label{eq:oscil_qp1} \\
    \delta\hat{\alpha}_\mu(t) &= \eta\sum_\nu\hat{\alpha}_\nu^\dag\left(X_{\nu\mu}(\omega)e^{-i\omega t}+Y_{\nu\mu}^*(\omega)e^{i\omega t}\right). \label{eq:oscil_qp2}
\end{align}
Here $E_\mu$ is the $\mu^\textrm{th}$ quasiparticle energy.
$X_{\nu\mu}(\omega)$ and $Y_{\nu\mu}(\omega)$ are the forward and backward amplitudes, respectively.
The TDHFB Hamiltonian $\hat{H}(t)$ can be written as
\begin{equation}\label{eq:TDHFB}
    \hat{H}(t) = \hat{H}_0 + \delta\hat{H}(t),
\end{equation}
where 
\begin{equation}\label{eq:HFB}
    \hat{H}_0 = \sum_\mu E_\mu\hat{\alpha}_\mu^\dag\hat{\alpha}_\mu
\end{equation}
and the induced Hamiltonian is
\begin{equation}\label{eq:dHt}
    \delta\hat{H}(t) = \eta \left(\delta\hat{H}(\omega)e^{-i\omega t} + \delta\hat{H}^\dag(\omega)e^{i\omega t}\right)
\end{equation}
with
\begin{equation}\label{eq:dHw}
    \delta\hat{H}(\omega) = \frac{1}{2}\sum_{\mu\nu}\left(\delta H_{\mu\nu}^{20}(\omega)\hat{\alpha}_\mu^\dag\hat{\alpha}_\nu^\dag + \delta H_{\mu\nu}^{02}(\omega)\hat{\alpha}_\mu\hat{\alpha}_\nu\right)
\end{equation}
representing a small-amplitude oscillation, and expressions for $\delta H_{\mu\nu}^{20}(\omega)$ and $\delta H_{\mu\nu}^{02}(\omega)$ given, e.g., in Ref.\ \cite{ring1983nuclear}.
Substituting these definitions into the TDHFB equation~\eqref{eq:TDHFB_eq} yields the FAM equations \cite{avogadro2011finite,mustonen2014finite}:
\begin{align}
    \left(E_\mu+E_\nu-\omega\right)X_{\mu\nu}(\omega)+\delta H_{\mu\nu}^{20}(\omega) &= -F_{\mu\nu}^{20}(\omega), \label{eq:lrp1} \\
    \left(E_\mu+E_\nu+\omega\right)Y_{\mu\nu}(\omega)+\delta H_{\mu\nu}^{02}(\omega) &= -F_{\mu\nu}^{02}(\omega). \label{eq:lrp2}
\end{align}
The quantity of interest is the \textit{strength function}, defined by
\begin{equation}\label{eq:strength}
    \frac{dB(\omega;\hat{F})}{d\omega} = -\frac{1}{\pi}\text{Im}\left\{S(\omega;\hat{F})\right\},
\end{equation}
where
\begin{equation}\label{eq:response_amplitude}
    S(\omega;\hat{F}) = \sum_{\mu<\nu}\left(F_{\mu\nu}^{20}(\omega)^*X_{\mu\nu}(\omega) + F_{\mu\nu}^{02}(\omega)^*Y_{\mu\nu}(\omega)\right).
\end{equation}
We refer to $S(\omega;\hat{F})$ in Eq.~\eqref{eq:response_amplitude} as the QRPA response function for the external field $\hat{F}$; it is the auxiliary quantity from which the strength function is obtained.
In practice, we assume that the external weak field is frequency-independent, i.e., $F_{\mu\nu}^{20}(\omega)=F_{\mu\nu}^{20}$ and $F_{\mu\nu}^{02}(\omega)=F_{\mu\nu}^{02}$.

\subsection{Obtaining
the Strength Function by Solving Linear Equations}

The FAM equations can also be written in a compact matrix form. This reformulation is useful for two reasons. First, it isolates the frequency dependence. Second, it makes it possible to apply standard linear-algebra tools directly to the QRPA strength-function problem.
Expanding the induced Hamiltonian matrix elements $\delta H_{\mu\nu}^{20}(\omega)$ and $\delta H_{\mu\nu}^{02}(\omega)$ in terms of the QRPA matrices $A$ and $B$ \cite{ring1983nuclear} via 
\begin{align}
    \delta H_{\mu\nu}^{20}(\omega) &= -(E_\mu+E_\nu)X_{\mu\nu}(\omega) + \sum_{\mu'<\nu'}\left(A_{\mu\nu,\mu'\nu'}X_{\mu'\nu'}(\omega)+B_{\mu\nu,\mu'\nu'}Y_{\mu'\nu'}(\omega)\right), \label{eq:dHw_expan1} \\
    \delta H_{\mu\nu}^{02}(\omega) &= -(E_\mu+E_\nu)Y_{\mu\nu}(\omega) + \sum_{\mu'<\nu'}\left(B_{\mu\nu,\mu'\nu'}^*X_{\mu'\nu'}(\omega)+A_{\mu\nu,\mu'\nu'}^*Y_{\mu'\nu'}(\omega)\right) \label{eq:dHw_expan2} \,,
\end{align}
we recast the FAM equations into the linear response matrix equation:
\begin{equation}\label{eq:lre}
    \left(
    \begin{bmatrix}
        A & B \\
        B^* & A^*
    \end{bmatrix}
    - \omega 
    \begin{bmatrix}
        I & 0 \\
        0 & -I
    \end{bmatrix}
    \right)
    \begin{bmatrix}
        X(\omega) \\
        Y(\omega)
    \end{bmatrix} = 
    - \begin{bmatrix}
        F^{20} \\
        F^{02}
    \end{bmatrix} \,,
\end{equation}
where $X(\omega),Y(\omega),F^{20},F^{02}\in\mathbb{C}^n$ are the vectorized representations of the strict upper triangular parts $(\mu<\nu)$ of the respective matrices.
One can evaluate the strength function on a frequency grid, $\omega_i$, $i=1,2,\ldots,n_\omega$, within an appropriate frequency range by solving \eqref{eq:lre} and using \eqref{eq:strength} and \eqref{eq:response_amplitude} for each $\omega_i$.  
Note that, when $\omega_i$ is near one of the eigenvalues of the matrix pencil
\begin{equation}
\left(\begin{bmatrix}
    A & B \\
    B^* & A^*
\end{bmatrix},
\begin{bmatrix}
    I & 0 \\
    0 & -I 
\end{bmatrix}
\right) \,,
\label{eq:matrixpen}
\end{equation}
\eqref{eq:lre} becomes ill-conditioned or singular.
To avoid these singularities in practical calculations, we replace the real frequency by the complex value $\omega \to \omega_\gamma := \omega + i\gamma$ with a smearing width $\gamma > 0$. 
The strength function then becomes,
\begin{equation}\label{eq:strength2}
    \frac{dB(\omega;\hat{F})}{d\omega} = \lim_{\gamma\to0^+}-\frac{1}{\pi}\text{Im}\left\{S(\omega_\gamma;\hat{F})\right\} = \lim_{\gamma\to0^+}-\frac{1}{\pi}\text{Im}\left\{\begin{bmatrix}
        F^{20} \\
        F^{02}
    \end{bmatrix}^\dag
    \begin{bmatrix}
        X(\omega_\gamma) \\
        Y(\omega_\gamma)
    \end{bmatrix}\right\} \,.
\end{equation}
\subsection{Representing
the Strength Function through Eigendecomposition}

An alternative way to compute the strength function is to first perform an eigendecomposition of the QRPA matrix pencil \eqref{eq:matrixpen}, and then rewrite the strength function in terms of the QRPA eigenmodes. 
This approach allows us to evaluate the strength function at any frequency $\omega$ once the eigendecomposition of \eqref{eq:matrixpen} is obtained. The special matrix structure of \eqref{eq:matrixpen} yields a corresponding structure in its eigendecomposition, as described below.

If the QRPA matrices satisfy $A^\dag=A, B^T=B$ and the matrix $\begin{bmatrix}
    A & B \\
    B^* & A^*
\end{bmatrix}$ is positive-definite, there exists an eigendecomposition whose eigenvalues are real and come in pairs.
We refer the reader to \cite[Proposition 1]{bjelvcic2022chebyshev} and \cite[Theorem 3]{shao2016structure} for a proof.
\begin{theorem}\label{thm:eigen}
    Let $A,B\in\mathbb{C}^{n\times n}$ such that $A^\dag=A,B^T=B$ and $\begin{bmatrix}
    A & B \\
    B^* & A^*
\end{bmatrix}$ is positive-definite.
Then there exist $X,Y\in\mathbb{C}^{n\times n}$ and diagonal $\Omega\in\mathbb{R}^{n\times n}$ with positive diagonal elements such that:
\begin{equation}\label{eq:eigendecomp}
    \begin{bmatrix}
        A & B \\
        B^* & A^*
    \end{bmatrix}
    \begin{bmatrix}
        X & Y^* \\
        Y & X^*
    \end{bmatrix} =
    \begin{bmatrix}
        I & 0 \\
        0 & -I
    \end{bmatrix}
    \begin{bmatrix}
        X & Y^* \\
        Y & X^*
    \end{bmatrix}
    \begin{bmatrix}
        +\Omega & 0 \\
        0 & -\Omega
    \end{bmatrix}\,,
\end{equation}
and
\begin{align}
    \begin{bmatrix}
        X & Y^* \\
        Y & X^*
    \end{bmatrix}^\dag
    \begin{bmatrix}
        I & 0 \\
        0 & -I
    \end{bmatrix}
    \begin{bmatrix}
        X & Y^* \\
        Y & X^*
    \end{bmatrix} =
    \begin{bmatrix}
        I & 0 \\
        0 & -I
    \end{bmatrix}, \label{eq:normalization1} \\
    \begin{bmatrix}
        X & Y^* \\
        Y & X^*
    \end{bmatrix}
    \begin{bmatrix}
        I & 0 \\
        0 & -I
    \end{bmatrix}
    \begin{bmatrix}
        X & Y^* \\
        Y & X^*
    \end{bmatrix}^\dag =
    \begin{bmatrix}
        I & 0 \\
        0 & -I
    \end{bmatrix}, \label{eq:normalization2} \\
    \begin{bmatrix}
        X & Y^* \\
        Y & X^*
    \end{bmatrix}^{-1} =
    \begin{bmatrix}
        I & 0 \\
        0 & -I
    \end{bmatrix}
    \begin{bmatrix}
        X & Y^* \\
        Y & X^*
    \end{bmatrix}^\dag
    \begin{bmatrix}
        I & 0 \\
        0 & -I
    \end{bmatrix}\,. \label{eq:normalization3}
\end{align}
\end{theorem}

By Theorem~\ref{thm:eigen}, the matrix in \eqref{eq:lre} is invertible for any $\omega_\gamma$ away from the real poles, and in particular for every $\gamma>0$:
\begin{equation}\label{eq:lre_inv}
\begin{split}
    &\left(\begin{bmatrix}
        A & B \\
        B^* & A^*
    \end{bmatrix}
    - \omega_\gamma 
    \begin{bmatrix}
        I & 0 \\
        0 & -I
    \end{bmatrix}
    \right)^{-1}
    = \begin{bmatrix}
        X & Y^* \\
        Y & X^*
    \end{bmatrix}
    \begin{bmatrix}
        (\Omega-\omega_\gamma I)^{-1} & 0 \\
        0 & (-\Omega-\omega_\gamma I)^{-1}
    \end{bmatrix}
    \begin{bmatrix}
        X & Y^* \\
        Y & X^*
    \end{bmatrix}^{-1}
    \begin{bmatrix}
        I & 0 \\
        0 & -I
    \end{bmatrix}\,.
\end{split}
\end{equation}
For compactness in the following derivation, we define
\begin{equation*}
    \mathcal{U}:=
    \begin{bmatrix}
        X & Y^* \\
        Y & X^*
    \end{bmatrix},
    \qquad
    \mathcal{F}:=
    \begin{bmatrix}
        F^{20} \\
        F^{02}
    \end{bmatrix}\,.
\end{equation*}
Using the vectorized notation introduced above and substituting the inverse~\eqref{eq:lre_inv} into Eq.~\eqref{eq:response_amplitude} yields
\begin{equation}\label{eq:strength3}
\begin{split}
    S(\omega_\gamma;\hat{F}) &= 
    \mathcal{F}^\dag
    \begin{bmatrix}
        X(\omega_\gamma) \\
        Y(\omega_\gamma)
    \end{bmatrix} \\
    &=
    -\mathcal{F}^\dag
    \left(\begin{bmatrix}
        A & B \\
        B^* & A^*
    \end{bmatrix}
    - \omega_\gamma 
    \begin{bmatrix}
        I & 0 \\
        0 & -I
    \end{bmatrix}
    \right)^{-1}
    \mathcal{F} \\
    &= -\left(\mathcal{U}^\dag\mathcal{F}\right)^\dag
    \begin{bmatrix}
        (\Omega-\omega_\gamma I)^{-1} & 0 \\
        0 & (-\Omega-\omega_\gamma I)^{-1}
    \end{bmatrix}
    \left(
    \mathcal{U}^{-1}
    \begin{bmatrix}
        I & 0 \\
        0 & -I
    \end{bmatrix}
    \mathcal{F}
    \right) \\
    &= -\left(\mathcal{U}^\dag\mathcal{F}\right)^\dag
    \begin{bmatrix}
        (\Omega-\omega_\gamma I)^{-1} & 0 \\
        0 & (\Omega+\omega_\gamma I)^{-1}
    \end{bmatrix}
    \left(\mathcal{U}^\dag\mathcal{F}\right)\,.
\end{split}
\end{equation}
This implies that
\begin{equation}\label{eq:Im_strength}
\begin{split}
    -\frac{1}{\pi}\text{Im}\left\{S(\omega_\gamma;\hat{F})\right\} = 
    \left(
    \begin{bmatrix}
        X & Y^* \\
        Y & X^*
    \end{bmatrix}^\dag
    \begin{bmatrix}
        F^{20} \\
        F^{02}
    \end{bmatrix}
    \right)^\dag
    \begin{bmatrix}
        \frac{\gamma/\pi}{(\omega-\Omega)^2+\gamma^2} & 0 \\
        0 & \frac{-\gamma/\pi}{(\omega+\Omega)^2+\gamma^2}
    \end{bmatrix}
    \left(
    \begin{bmatrix}
        X & Y^* \\
        Y & X^*
    \end{bmatrix}^\dag
    \begin{bmatrix}
        F^{20} \\
        F^{02}
    \end{bmatrix}
    \right)\,,
\end{split}
\end{equation}
where we have used the corresponding identity for the diagonal matrix appearing above.
It follows, after taking the limit $\gamma\to0^+$, that the strength function~\eqref{eq:strength2} can be expressed as a sum of Dirac delta functions located at the QRPA poles,
\begin{equation}\label{eq:response3}
    \frac{dB(\omega;\hat{F})}{d\omega} = \sum_{i=1}^n |T_i(\hat{F})|^2\delta(\omega-\Omega_i) - \sum_{i=1}^n |\tilde{T}_i(\hat{F})|^2\delta(\omega+\Omega_i),
\end{equation}
where $T_i(\hat{F})$ and $\tilde{T}_i(\hat{F})$ are defined as
\begin{equation}\label{eq:Ti}
    \begin{bmatrix}
        T(\hat{F}) \\
        \tilde{T}(\hat{F})
    \end{bmatrix}
    = \begin{bmatrix}
        X & Y^* \\
        Y & X^*
    \end{bmatrix}^\dag
    \begin{bmatrix}
        F^{20} \\
        F^{02}
    \end{bmatrix}.
\end{equation}
In the eigenmode representation of Eq.~\eqref{eq:response3}, the projected quantities $T_i(\hat{F})$ and $\tilde{T}_i(\hat{F})$ are often referred to as QRPA transition amplitudes to the positive- and negative-frequency branches, respectively.

\begin{figure}[htbp]
\centering
\includegraphics[scale=0.1]{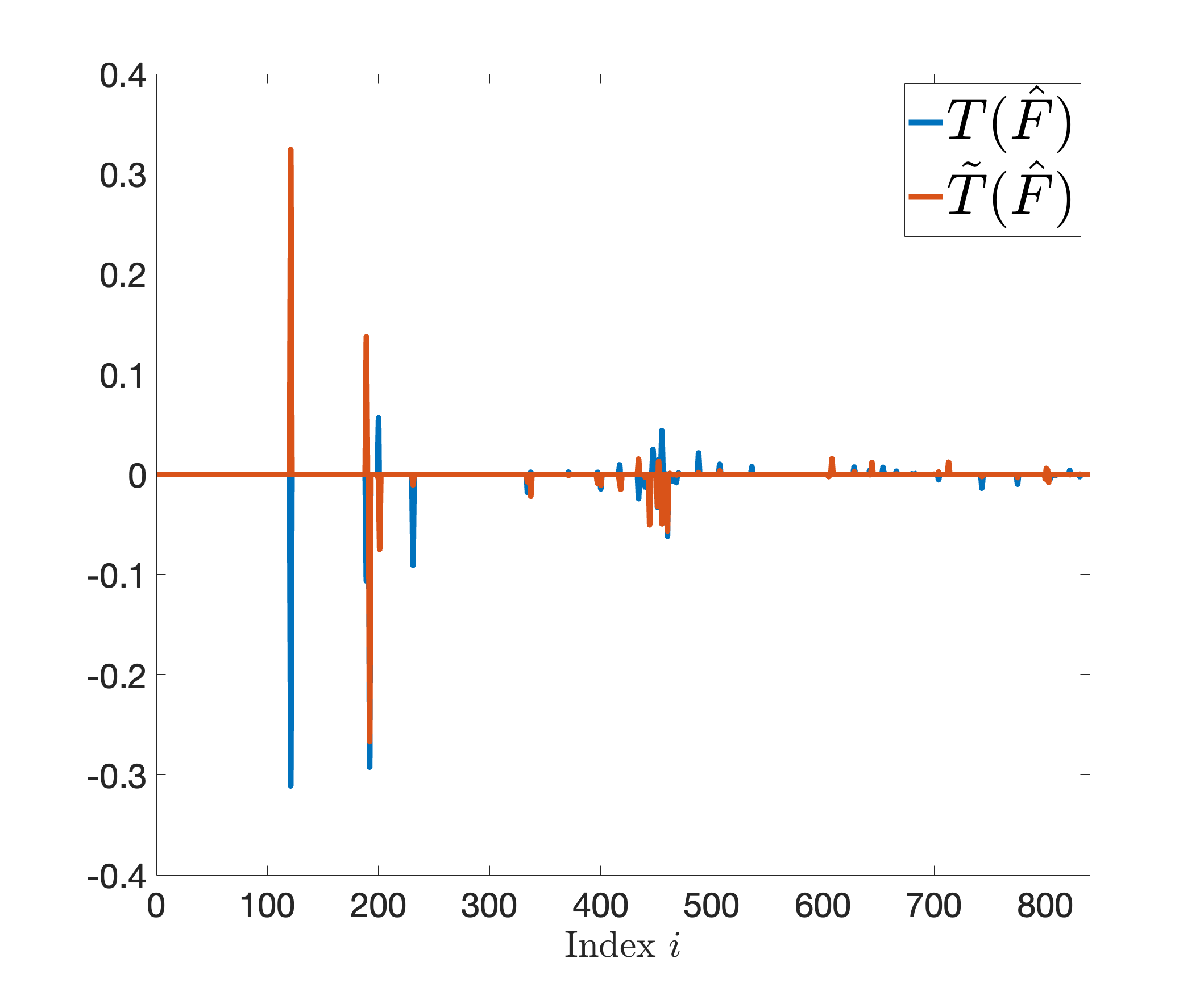}
\includegraphics[scale=0.1]{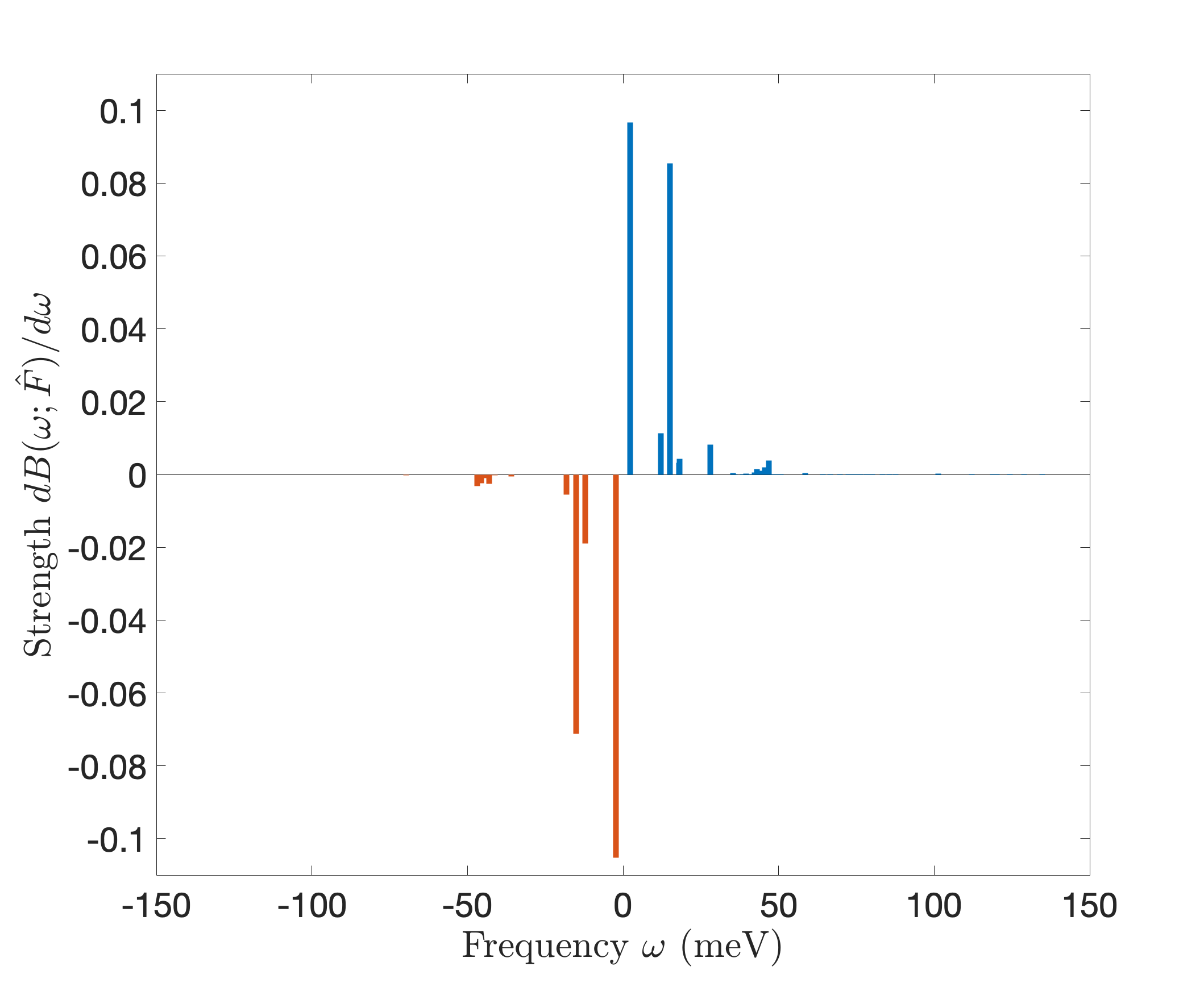}
\caption{\label{fig:strength_transition} Illustrative example based on a five-shell QRPA calculation in $^{6}$Li.  The model space is small enough that the QRPA matrices can be formed and diagonalized explicitly. The example is used here only to visualize the spectral representation: left, representative transition amplitudes for the positive- and negative-frequency branches; right, the corresponding schematic strength function, shown as a sum of weighted discrete transitions.}
\end{figure}

In Figure~\ref{fig:strength_transition}, we plot example QRPA transition amplitudes and the corresponding strength function.
The magnitudes of $T_i(\hat{F})$ and $\tilde{T}_i(\hat{F})$ do not necessarily match, so the resulting strength function is not, in general, antisymmetric about the origin.

\subsection{Reformulation as a Reduced Eigenvalue Problem}

In practice, the QRPA matrices $A$ and $B$ and the external weak fields $F^{20},F^{02}$ are often real.
In that case, the dimension of the eigenvalue problem can be reduced by two, by using the products $MK$ and $KM$, where $M=A+B$ and $K=A-B$. Under the positivity assumptions inherited from Theorem~\ref{thm:eigen}, both $M$ and $K$ are symmetric positive-definite. Related reductions also appear in \cite{ring1983nuclear,brabec2015efficient}. We summarize this construction here for completeness.

The real counterpart of the eigendecomposition~\eqref{eq:eigendecomp} is
\begin{equation}\label{eq:eigendecomp_real}
    \begin{bmatrix}
        A & B \\
        B & A
    \end{bmatrix}
    \begin{bmatrix}
        X & Y \\
        Y & X
    \end{bmatrix} =
    \begin{bmatrix}
        I & 0 \\
        0 & -I
    \end{bmatrix}
    \begin{bmatrix}
        X & Y \\
        Y & X
    \end{bmatrix}
    \begin{bmatrix}
        +\Omega & 0 \\
        0 & -\Omega
    \end{bmatrix},
\end{equation}
with $X,Y\in\mathbb{R}^{n\times n}$.
In addition, the normalization constraints~\eqref{eq:normalization1},\eqref{eq:normalization2} indicate that
\begin{equation}\label{eq:normalization4}
    X^TX-Y^TY = I, \quad X^TY-Y^TX=0, \quad XX^T-YY^T=I, \quad XY^T-YX^T = 0.
\end{equation}
These conditions imply that
\begin{equation}\label{eq:normalization5}
    (X+Y)^{-1} = (X-Y)^T \quad\text{and}\quad (X-Y)^{-1} = (X+Y)^T.
\end{equation}
Note that the eigendecomposition~\eqref{eq:eigendecomp_real} is equivalent to
\begin{equation}\label{eq:eigendecomp_real2}
    \begin{bmatrix}
        A & B \\
        -B & -A
    \end{bmatrix}
    \begin{bmatrix}
        X & Y \\
        Y & X
    \end{bmatrix} =
    \begin{bmatrix}
        X & Y \\
        Y & X
    \end{bmatrix}
    \begin{bmatrix}
        +\Omega & 0 \\
        0 & -\Omega
    \end{bmatrix},
\end{equation}
and the matrix 
\begin{equation}\label{eq:Casida}
    \begin{bmatrix}
        A & B \\
        -B & -A
    \end{bmatrix}
\end{equation}
is often referred to as the Casida matrix.

One can rewrite the eigenvalue problem for the Casida matrix,
\begin{equation}\label{eq:evp}
    \begin{bmatrix}
        A & B \\
        -B & -A
    \end{bmatrix}
    \begin{bmatrix}
        x_i \\
        y_i
    \end{bmatrix} =
    \Omega_i
    \begin{bmatrix}
        x_i \\
        y_i
    \end{bmatrix},
\end{equation}
by employing a unitary similarity transformation,
\begin{equation}\label{eq:J}
    J = \frac{1}{\sqrt{2}}
    \begin{bmatrix}
        I & I \\
        I & -I
    \end{bmatrix}
\end{equation}
to obtain
\begin{equation}\label{eq:tran_evp}
    \begin{bmatrix}
        0 & K \\
        M & 0
    \end{bmatrix}
    \begin{bmatrix}
        x_i+y_i \\
        x_i-y_i
    \end{bmatrix}
    = \Omega_i
    \begin{bmatrix}
        x_i+y_i \\
        x_i-y_i
    \end{bmatrix},
\end{equation}
where $K:=A-B$ and $M:=A+B$.
Then we have
\begin{equation}\label{eq:evp_KM}
\begin{split}
     K(x_i-y_i) = \Omega_i (x_i+y_i), \\
     M(x_i+y_i) = \Omega_i (x_i-y_i),
\end{split}
\end{equation}
and
\begin{equation}\label{eq:evp_KM2}
\begin{split}
     MK(x_i-y_i) = \Omega_i^2 (x_i-y_i), \\
     KM(x_i+y_i) = \Omega_i^2 (x_i+y_i).
\end{split}
\end{equation}
In other words,
\begin{equation}\label{eq:MK_decomp}
    M = (X-Y)\Omega(X-Y)^T \quad\text{and}\quad K = (X+Y)\Omega(X+Y)^T,
\end{equation}
and 
\begin{align}
    MK &= (X-Y)\Omega^2(X-Y)^{-1}, \label{eq:MK} \\
    KM &= (X+Y)\Omega^2(X+Y)^{-1}, \label{eq:KM}
\end{align}

These identities show that the eigendecomposition~\eqref{eq:eigendecomp_real2} of dimension $2n$ can be replaced by an eigenvalue problem of dimension $n$ for either the matrix $MK$ or the matrix $KM$.
In addition, once we have computed an eigenvector $x_i-y_i$ of $MK$, we can recover $x_i+y_i$ from
\begin{equation}\label{eq:xipyi}
    x_i+y_i = \frac{1}{\Omega_i}K(x_i-y_i) \,,
\end{equation}
without having to find an eigendecomposition of $KM$.
Once the vectors $x_i+y_i$ and $x_i-y_i$ are known, the original QRPA eigenvectors $x_i$ and $y_i$ follow immediately.

The reduced formulation also gives a direct spectral representation of the strength function. To see this, we insert the unitary similarity matrix $J$~\eqref{eq:J} into
\begin{equation*}
    \begin{bmatrix}
        X & Y \\
        Y & X
    \end{bmatrix}^T
    \begin{bmatrix}
        F^{20} \\
        F^{02}
    \end{bmatrix}
\end{equation*}
as
\begin{equation*}
    \begin{bmatrix}
        X & Y \\
        Y & X
    \end{bmatrix}^T
    JJ
    \begin{bmatrix}
        F^{20} \\
        F^{02}
    \end{bmatrix}.
\end{equation*}
This yields
\begin{equation}\label{eq:Ri}
    \begin{bmatrix}
        R(\hat{F}) \\
        \tilde{R}(\hat{F})
    \end{bmatrix} :=
    \frac{1}{2}
    \begin{bmatrix}
        (X+Y)^T(F^{20}+F^{02}) + (X-Y)^T(F^{20}-F^{02}) \\
        (X+Y)^T(F^{20}+F^{02}) - (X-Y)^T(F^{20}-F^{02})
    \end{bmatrix},
\end{equation}
so that the strength function can be written as
\begin{equation}\label{eq:strength_Lanczos}
    \frac{dB(\omega;\hat{F})}{d\omega} = \sum_{i=1}^n |R_i(\hat{F})|^2\delta(\omega-\Omega_i) - \sum_{i=1}^n |\tilde{R}_i(\hat{F})|^2\delta(\omega+\Omega_i).
\end{equation}
This representation will serve as the starting point for the Lanczos approximation developed in the next section.

The reduced formulation above is the key structural ingredient for the numerical methods developed in the next section. In particular, it provides the reduced operator on which the symmetric Lanczos approximation is built.

\section{Computational Methods for the Strength Function}\label{sec:computational_methods}

This section describes the numerical strategies used to evaluate the QRPA strength function in practice. We first summarize two frequency-by-frequency approaches based on the linear response equation and then derive the symmetric Lanczos approximation based on the reduced eigenvalue problem.

\subsection{Frequency-by-Frequency Solvers}

\subsubsection{Finite Amplitude Method (FAM)}\label{subsec:FAM}

A standard approach to solving Eqs.~\eqref{eq:lrp1} and \eqref{eq:lrp2} is the iterative finite amplitude method (FAM) \cite{avogadro2011finite,mustonen2014finite}.
For each fixed frequency $\omega$, FAM treats the response equations as a fixed-point problem for the amplitudes $X_{\mu\nu}(\omega)$ and $Y_{\mu\nu}(\omega)$.
A basic iteration proceeds as follows:
\begin{enumerate}
    \item Use the current amplitudes $X_{\mu\nu}(\omega)$ and $Y_{\mu\nu}(\omega)$ to compute $\delta H_{\mu\nu}^{20}(\omega)$ and $\delta H_{\mu\nu}^{02}(\omega)$.
    \item Add the external field contributions,
    \begin{align*}
        \delta H_{\mu\nu}^{20}(\omega) &\leftarrow \delta H_{\mu\nu}^{20}(\omega) + F_{\mu\nu}^{20},  \\
        \delta H_{\mu\nu}^{02}(\omega) &\leftarrow \delta H_{\mu\nu}^{02}(\omega) + F_{\mu\nu}^{02}.
    \end{align*}
    \item Update the amplitudes according to
    \begin{equation}\label{eq:FAM_updates}
        \begin{split}
            X_{\mu\nu}(\omega) &\leftarrow -\left(E_\mu+E_\nu-\omega\right)^{-1}\delta H_{\mu\nu}^{20}(\omega), \\
            Y_{\mu\nu}(\omega) &\leftarrow -\left(E_\mu+E_\nu+\omega\right)^{-1}\delta H_{\mu\nu}^{02}(\omega).
        \end{split}
    \end{equation}
\end{enumerate}
The FAM is effective because it avoids constructing the full QRPA matrix explicitly.
However, the plain fixed-point iteration is not guaranteed to converge for all frequencies: as with any stationary fixed-point scheme, convergence requires the associated iteration map to be sufficiently contractive.
In practice, convergence can become slow or fail near poorly conditioned response points, especially close to QRPA poles.
For this reason, practical FAM calculations commonly use acceleration schemes such as Broyden mixing~\cite{broyden1965class}.
Even with such acceleration, the full iteration must be repeated at every frequency point in the target interval.

\subsubsection{GMRES Solver}

Because Eq.~\eqref{eq:lre} is a linear system, Krylov-subspace methods such as the generalized minimal residual method (GMRES) \cite{saad1986gmres} provide a natural alternative to fixed-point FAM iterations. For a given complex frequency $\omega_\gamma$, GMRES seeks the approximation in the associated Krylov subspace that minimizes the residual norm. In our calculations, this typically leads to significantly fewer iterations than plain FAM updates.

To build the Krylov basis, the matrices $A$ and $B$ need not be formed explicitly. Each GMRES iteration only requires the matrix--vector product
\begin{equation}\label{eq:lre_matvec}
    \begin{bmatrix}
    X(\omega_\gamma) \\ Y(\omega_\gamma)
\end{bmatrix} \to
\left(
    \begin{bmatrix}
        A & B \\
        B & A
    \end{bmatrix}
    - \omega_\gamma 
    \begin{bmatrix}
        I & 0 \\
        0 & -I
    \end{bmatrix}
    \right)
    \begin{bmatrix}
        X(\omega_\gamma) \\
        Y(\omega_\gamma)
    \end{bmatrix}.
\end{equation}
Evaluating this product is equivalent to computing the left-hand side of the FAM equations~\eqref{eq:lrp1} and \eqref{eq:lrp2}. As with the FAM, however, a separate GMRES solve is required for every frequency value of interest. In Section~\ref{subsec:matvecs}, we describe how this matvec is assembled in practice and present the preconditioner used to accelerate GMRES convergence.

\subsection{Lanczos Approximation of the Strength Function}

We now use the reduced formulation to derive a Lanczos approximation for the strength function. Instead of fully diagonalizing~\eqref{eq:MK} (or \eqref{eq:KM}), which can still be expensive even after reduction, we approximate the exact spectral representation~\eqref{eq:strength_Lanczos} by means of the Lanczos method \cite{lanczos1950iteration}.
From a numerical perspective, the appeal of this approach is that one Lanczos run can capture the dominant spectral information needed over a broad frequency interval.
Closely related Krylov approaches have also been used for linear-response strength calculations in nuclear systems; see, for example, \cite{toivanen2010linear}.

Although the matrix $MK$ is not symmetric in the Euclidean inner product, it is self-adjoint with respect to the $K$-inner product, i.e.,
\begin{equation}\label{eq:Kinner}
    \langle x, MKy\rangle_K = x^TKMKy = (MKx)^TKy = \langle MKx, y\rangle_K,
\end{equation}
for any $x,y\in\mathbb{R}^n$.
This observation allows us to apply a symmetric Lanczos process in the $K$-inner product rather than working with a nonsymmetric algorithm in the Euclidean inner product.
An $m$-step Lanczos process generates
\begin{equation}\label{eq:Lanczos}
    MKQ_m = Q_mT_m+f_me_m^T
\end{equation}
where
\begin{equation}\label{eq:Qm_Lanczos}
    Q_m^TKQ_m = I \quad\text{and}\quad Q_m^TKf_m = 0.
\end{equation}
The projected matrix $T_m\in\mathbb{R}^{m\times m}$ is tridiagonal and has the eigendecomposition
\begin{equation*}
    T_mY_m = Y_m\Theta_m.
\end{equation*}
The eigenvalues $\{\theta_i\}_{i=1}^m$ of the tridiagonal matrix $T_m$ approximate the squared excitation energies $\{\Omega_i^2\}_{i=1}^n$. The corresponding excitation energies are therefore approximated as
\begin{equation}\label{eq:eigs_appr}
    \Omega_i\approx\sqrt{\theta_i}.
\end{equation}
The eigenvectors $(X-Y)\in\mathbb{R}^{n\times n}$ of $MK$ are approximated by $Z_m\in\mathbb{R}^{n\times m}$, defined as
\begin{equation}\label{eq:MK_Ritzvecs}
    Z_m:=Q_mY_m.
\end{equation}
These Ritz vectors, however, are normalized through the relation
\begin{equation}\label{eq:normal1}
    Z_m^TKZ_m=I,
\end{equation}
which is different from the normalization
\begin{equation}\label{eq:reduced_normalization}
    (X-Y)^TK(X-Y)=\Omega
\end{equation}
that holds for the eigenvector $X-Y$ of $MK$ as in \eqref{eq:MK_decomp}.

To relate this normalization to the exact QRPA normalization, suppose for a moment that the reduced eigenproblem were solved exactly (for example, through an $n$-step Lanczos process) and that we had obtained
\begin{equation}\label{eq:MK_evp}
    MKZ=Z\Theta
\end{equation}
where
\begin{equation}\label{eq:normal2}
    Z^TKZ=I \quad \mbox{and} \quad \Theta = \Omega^2.
\end{equation}
Comparing the normalization~\eqref{eq:normal2} with the normalization~\eqref{eq:reduced_normalization}, we immediately obtain
\begin{equation}\label{eq:X-minus-Y}
    X-Y=Z\Omega^{1/2} = Z\Theta^{1/4}.
\end{equation}
Moreover, since
\begin{equation*}
    X+Y=K(X-Y)\Omega^{-1},
\end{equation*}
we have
\begin{equation}\label{eq:X-plus-Y}
    X+Y = KZ\Omega^{-1/2} = KZ\Theta^{-1/4}.
\end{equation}

Thus, once the Ritz pairs $(\Theta_m,Z_m)$ have been computed from an $m$-step Lanczos run, the strength function is approximated by
\begin{equation}\label{eq:lanczos_strength_approx}
    \frac{dB(\omega;\hat{F})}{d\omega} \approx \sum_{i=1}^m |S_i(\hat{F})|^2\delta(\omega-\sqrt{\theta_i}) - \sum_{i=1}^m |\tilde{S}_i(\hat{F})|^2\delta(\omega+\sqrt{\theta_i}),
\end{equation}
where the vectors $S(\hat{F})$ and $\tilde{S}(\hat{F})$ have components $S_i(\hat{F})$ and $\tilde{S}_i(\hat{F})$, respectively, and are defined as
\begin{equation}\label{eq:Si}
    \begin{bmatrix}
        S(\hat{F}) \\
        \tilde{S}(\hat{F})
    \end{bmatrix}
    = \frac{1}{2}
    \begin{bmatrix}
        (KZ_m\Theta_m^{-1/4})^T(F^{20}+F^{02}) + (Z_m\Theta_m^{1/4})^T(F^{20}-F^{02}) \\
        (KZ_m\Theta_m^{-1/4})^T(F^{20}+F^{02}) - (Z_m\Theta_m^{1/4})^T(F^{20}-F^{02})
    \end{bmatrix}.
\end{equation}

One might worry that replacing the full $n$-term spectral sum by only $m$ Ritz contributions is too severe an approximation, or that the apparent agreement is driven primarily by replacing each Dirac delta peak by a smooth kernel such as a Lorentzian. The smoothing is certainly important for comparing strength profiles on a finite-frequency grid, but the quality of the approximation also depends on whether the projected transition amplitudes $S_i(\hat F)$ and $\tilde S_i(\hat F)$ capture the relevant QRPA strength carried by the external field. In the numerical section of this paper, we therefore assess the smoothed strength functions through pointwise agreement, KL divergence, and runtime comparisons.

\subsubsection{Comparison with Frequency-by-Frequency Solvers}
The conceptual difference between the Lanczos approach and the FAM/GMRES solvers is worth emphasizing. The Lanczos method approximates the spectral representation of the strength function, whereas the FAM and GMRES approximate the response function at one prescribed frequency.
Note that the strength function defined as in~\eqref{eq:strength},
\begin{equation*}
    \frac{dB(\omega;\hat{F})}{d\omega} = -\frac{1}{\pi}\text{Im}\left\{S(\omega;\hat{F})\right\},
\end{equation*}
where
\begin{equation*}
    S(\omega;\hat{F}) = \sum_{\mu<\nu}\left(F_{\mu\nu}^{20}(\omega)^*X_{\mu\nu}(\omega) + F_{\mu\nu}^{02}(\omega)^*Y_{\mu\nu}(\omega)\right),
\end{equation*}
requires the amplitudes $X_{\mu\nu}(\omega)$ and $Y_{\mu\nu}(\omega)$ for a \textit{fixed} frequency $\omega$.
Consequently, both the FAM and GMRES must be run separately for each frequency at which the strength function is requested.

On the other hand, the strength function written in terms of the eigendecomposition as in~\eqref{eq:response3},
\begin{equation*}
    \frac{dB(\omega;\hat{F})}{d\omega} = \sum_{i=1}^n |T_i(\hat{F})|^2\delta(\omega-\Omega_i) - \sum_{i=1}^n |\tilde{T}_i(\hat{F})|^2\delta(\omega+\Omega_i),
\end{equation*}
where 
\begin{equation*}
    \begin{bmatrix}
        T(\hat{F}) \\
        \tilde{T}(\hat{F})
    \end{bmatrix}
    = \begin{bmatrix}
        X & Y^* \\
        Y & X^*
    \end{bmatrix}^\dag
    \begin{bmatrix}
        F^{20} \\
        F^{02}
    \end{bmatrix},
\end{equation*}
shows that $\omega$ is decoupled from the transition-amplitude calculation. Because the Lanczos procedure approximates the eigenvalues $\Omega_i$ together with the QRPA transition amplitudes $T(\hat{F})$ and $\tilde{T}(\hat{F})$, a single Lanczos run can be used to approximate the strength function over an entire frequency interval. The FAM and GMRES are therefore attractive when only a small number of frequencies are needed, whereas the Lanczos method becomes advantageous when one seeks a broad spectral scan or a full response profile.

\subsubsection{Lanczos Initialization}

The symmetric Lanczos process is applied to the reduced matrix $MK$ with the $K$-inner product. We initialize the recursion with the normalized vector associated with the external field combination
\begin{equation}
    f_{+} := F^{20}+F^{02}, \qquad q_1 = \frac{f_{+}}{\|f_{+}\|_{K}}, \qquad \|v\|_{K}:=(v^TKv)^{1/2}.
\end{equation}
This choice is natural in the reduced formulation because the combination $X+Y$, which is reconstructed from $KZ_m\Theta_m^{-1/4}$, couples directly to $f_{+}$ in Eq.~\eqref{eq:Si}. It also respects the symmetry structure of the axially-deformed calculation. In our implementation, the quasiparticle basis and QRPA matrices inherit selection rules associated with the intrinsic projection quantum number, so $A$, $B$, and the reduced operator $MK$ decompose into invariant symmetry sectors. Because $F^{20}$ and $F^{02}$ are generated from an external transition operator with a specified intrinsic component, $f_{+}$ already lies in the sector relevant to the desired response.

This symmetry-adapted support is important in practice. An unrestricted random starting vector generally contains components in other decoupled sectors and therefore samples spectral information unrelated to the chosen strength function. A random vector restricted to the same nonzero pattern and sign structure as $f_{+}$ removes most of this mismatch and can reproduce the overall profile well, but it still lacks the physical matrix-element magnitudes of the external field. Starting from $f_{+}$ therefore both selects the correct symmetry sector and preserves the field-dependent weights needed for the detailed strength profile. We also retain the projections associated with $f_-:=F^{20}-F^{02}$ during the Lanczos calculation, since Eq.~\eqref{eq:Si} shows that both $f_{+}$ and $f_-$ enter the reconstructed transition amplitudes.

\section{Numerical Results}\label{sec:numerical_results}

In this section we assess the performance of the symmetric Lanczos method for approximating QRPA strength functions.
We first discuss several practical considerations, including the regularization of the Dirac-delta distribution and the construction of matrix--vector products (matvecs) for GMRES and Lanczos.
We then present calculations for two realistic nuclei: the medium-mass nucleus $^{112}$Sn and the heavier rare-earth nucleus $^{150}$Nd.
The numerical discussion is organized in stages: first, we compare GMRES with the conventional iterative FAM (IFAM) to establish GMRES as an efficient frequency-by-frequency reference calculation; next, we compare the Lanczos approximation directly with this GMRES reference using strength profiles, pointwise errors, KL divergences, and runtimes.

All iterative solvers used in this section were implemented through the Python package \texttt{pynfam} \cite{ney2020global}.
This package provides a workflow around the Fortran codes \texttt{hfbtho} and \texttt{pnfam} for charge-changing HFB+QRPA calculations \cite{marevic2022hfbtho,mustonen2014finite}.
For each nucleus, \texttt{hfbtho} is first used to obtain the HFB ground state through an axial deformation scan, after which the lowest-energy solution is taken as the reference state for the subsequent \texttt{pnfam} calculation.
The latter computes the linear response for the chosen external field on the prescribed energy contour, from which the corresponding strength function is constructed.

Unless otherwise noted, we use common numerical settings across the two nuclei and vary only the nucleus-dependent inputs; both $^{112}$Sn and $^{150}$Nd are computed with 16 harmonic-oscillator shells.
The underlying mean field is generated with SkM$^*$, a standard Skyrme energy-density functional widely used in self-consistent mean-field and QRPA calculations \cite{bartel1982skm}.
In the pnFAM calculations we do not explicitly include the $J^2$ terms; that is, the spin-current contributions associated with that sector of the Skyrme functional are not turned on by hand in our setup.
Detailed discussions of the individual nuclei and their corresponding response functions are deferred to the subsections below.

\subsection{Practical Considerations}

\subsubsection{Regularization}

Because no conventional function possesses the properties of the Dirac-delta function, it is often defined through limits or the theory of distributions.
A common way to define the delta function is as the limit of a Gaussian function
\begin{equation}\label{eq:Gaussian}
    g_\sigma(t) = \frac{1}{(2\pi\sigma^2)^{1/2}} e^{-\frac{t^2}{2\sigma^2}},
\end{equation}
or as the limit of a Lorentzian function
\begin{equation}\label{eq:Lorentzian}
    L_\gamma(t) = \frac{\gamma/\pi}{t^2+\gamma^2}.
\end{equation}
As the width parameters $\sigma$ and $\gamma$ of the Gaussian and Lorentzian, respectively, approach zero, the corresponding functions become infinitely tall and infinitely narrow, resembling the Dirac-delta function.

For density-of-states (DOS) problems, replacing the delta function with such an approximate function is known as \textit{regularizing} the spectral density, and it is shown in \cite{lin2016approximating} that the width parameter controls the resolution of the DOS curves.
In this paper, we adapt this regularization strategy for our strength function calculations, replacing the delta functions with Lorentzian functions.

\subsubsection{Matvecs for GMRES and Lanczos}\label{subsec:matvecs}

The equivalence between the FAM equations~\eqref{eq:lrp1},\eqref{eq:lrp2} and the linear response equation~\eqref{eq:lre} indicates that we can obtain the matvec
\begin{equation}\label{eq:matvec}
    \begin{bmatrix}
    X(\omega_\gamma) \\ Y(\omega_\gamma)
    \end{bmatrix} \mapsto
    \left(
    \begin{bmatrix}
        A & B \\
        B & A
    \end{bmatrix}
    - \omega_\gamma 
    \begin{bmatrix}
        I & 0 \\
        0 & -I
    \end{bmatrix}
    \right)
    \begin{bmatrix}
        X(\omega_\gamma) \\
        Y(\omega_\gamma)
    \end{bmatrix}
\end{equation}
by performing the following steps for each $\mu<\nu$.
\begin{enumerate}
    \item Use the amplitudes $X_{\mu\nu}(\omega_\gamma)$ and $Y_{\mu\nu}(\omega_\gamma)$ to compute $\delta H_{\mu\nu}^{20}(\omega_\gamma)$ and $\delta H_{\mu\nu}^{02}(\omega_\gamma)$.
    \item Add $\delta H_{\mu\nu}^{20}(\omega_\gamma)$ to $\left(E_\mu+E_\nu-\omega_\gamma\right)X_{\mu\nu}(\omega_\gamma)$ and add $\delta H_{\mu\nu}^{02}(\omega_\gamma)$ to $\left(E_\mu+E_\nu+\omega_\gamma\right)Y_{\mu\nu}(\omega_\gamma)$.
\end{enumerate}

\paragraph{GMRES.}
For implementing GMRES, we use the diagonal preconditioner
\begin{equation}\label{eq:precond}
    P = \begin{bmatrix}
        \left(E_\mu+E_\nu-\omega_\gamma\right)^{-1} & 0 \\
        0 & \left(E_\mu+E_\nu+\omega_\gamma\right)^{-1}
    \end{bmatrix}.
\end{equation}
The corresponding preconditioned matvec is
\begin{equation}\label{eq:gmres_matvec}
    \begin{bmatrix}
        X(\omega_\gamma) \\ Y(\omega_\gamma)
    \end{bmatrix} \mapsto
    \begin{bmatrix}
        \left(E_\mu+E_\nu-\omega_\gamma\right)^{-1} & 0 \\
        0 & \left(E_\mu+E_\nu+\omega_\gamma\right)^{-1}
    \end{bmatrix}
    \left(\begin{bmatrix}
        A & B \\
        B & A
    \end{bmatrix}
    - \omega_\gamma 
    \begin{bmatrix}
        I & 0 \\
        0 & -I
    \end{bmatrix}\right)
    \begin{bmatrix}
        X(\omega_\gamma) \\
        Y(\omega_\gamma)
    \end{bmatrix},
\end{equation}
which can be obtained by performing the following steps:
\begin{enumerate}
    \item Use the amplitudes $X_{\mu\nu}(\omega_\gamma)$ and $Y_{\mu\nu}(\omega_\gamma)$ to compute $\delta H_{\mu\nu}^{20}(\omega_\gamma)$ and $\delta H_{\mu\nu}^{02}(\omega_\gamma)$.
    \item Add $\left(E_\mu+E_\nu-\omega_\gamma\right)^{-1}\delta H_{\mu\nu}^{20}(\omega_\gamma)$ to $X(\omega_\gamma)$ and add $\left(E_\mu+E_\nu+\omega_\gamma\right)^{-1}\delta H_{\mu\nu}^{02}(\omega_\gamma)$ to $Y(\omega_\gamma)$.
\end{enumerate}

\paragraph{Lanczos.}
In order to perform the matvec involving the matrix $MK$, where $M=A+B$ and $K=A-B$, for the symmetric Lanczos iterations on the reduced eigenvalue problem, we need to repeat the steps of \eqref{eq:matvec}. 
The crucial distinction in the Lanczos method is that we need to set $\omega_\gamma=0$, so that we have a matvec with the matrix
\[
\begin{bmatrix}
    A & B \\
    B & A
\end{bmatrix}.
\]
Given a vector $X\in\mathbb{R}^n$, the matvec with $MK$ is computed as follows:
\begin{enumerate}
    \item Set up a vector $\begin{bmatrix} X \\ 0 \end{bmatrix}$ by appending $X$ with a zero vector of size $n$.
    \item Apply the matvec to $\begin{bmatrix} X \\ 0 \end{bmatrix}$ and obtain $\begin{bmatrix} AX \\ BX \end{bmatrix}$.
    \item Compute $Z:=AX-BX$.
    \item Apply the matvec to $\begin{bmatrix} Z \\ 0 \end{bmatrix}$ and obtain $\begin{bmatrix} AZ \\ BZ \end{bmatrix}$.
    \item Set $MKX:=AZ+BZ$.
\end{enumerate}
We note that in addition to these two matvecs, we need one additional matvec to do $K$-inner product orthogonalization, another matvec for reorthogonalization, and a final one for normalization with respect to $K$-inner product.
Thus, the Lanczos recurrence itself incurs five matvecs per iteration.
In addition, after the $m$-step Lanczos iteration, we need to perform $m$ matvecs to compute $KZ_m\Theta_m^{-1/4}$ in order to derive the weights $S_i(\hat{F})$ and $\tilde{S}_i(\hat{F})$.
Therefore, $5m+m=6m$ matvecs are required in total.

\subsection{Medium-Mass and Heavy Nuclei}

The QRPA matrix dimensions are $94,482$ for the medium-mass nucleus $^{112}$Sn and $101,324$ for the heavier rare-earth nucleus $^{150}$Nd.
We use converged frequency-by-frequency solvers as references for the smeared strength functions.
We compute the Gamow--Teller strength by combining the intrinsic components of the $J^\pi=1^+$ spin--isospin operator.
In an axially-symmetric system, this amounts to performing separate calculations for the $K=0$ and $K=1$ components and reconstructing the total strength as the sum of the $K=0$ contribution and twice the $K=1$ contribution.
The factor of two accounts for the degeneracy of the $K=\pm1$ branches, which contribute equally to the total Gamow--Teller strength. 

\subsubsection{GMRES as a Frequency-by-Frequency Reference}

\begin{figure}[htbp]
\centering
\includegraphics[scale=0.1]{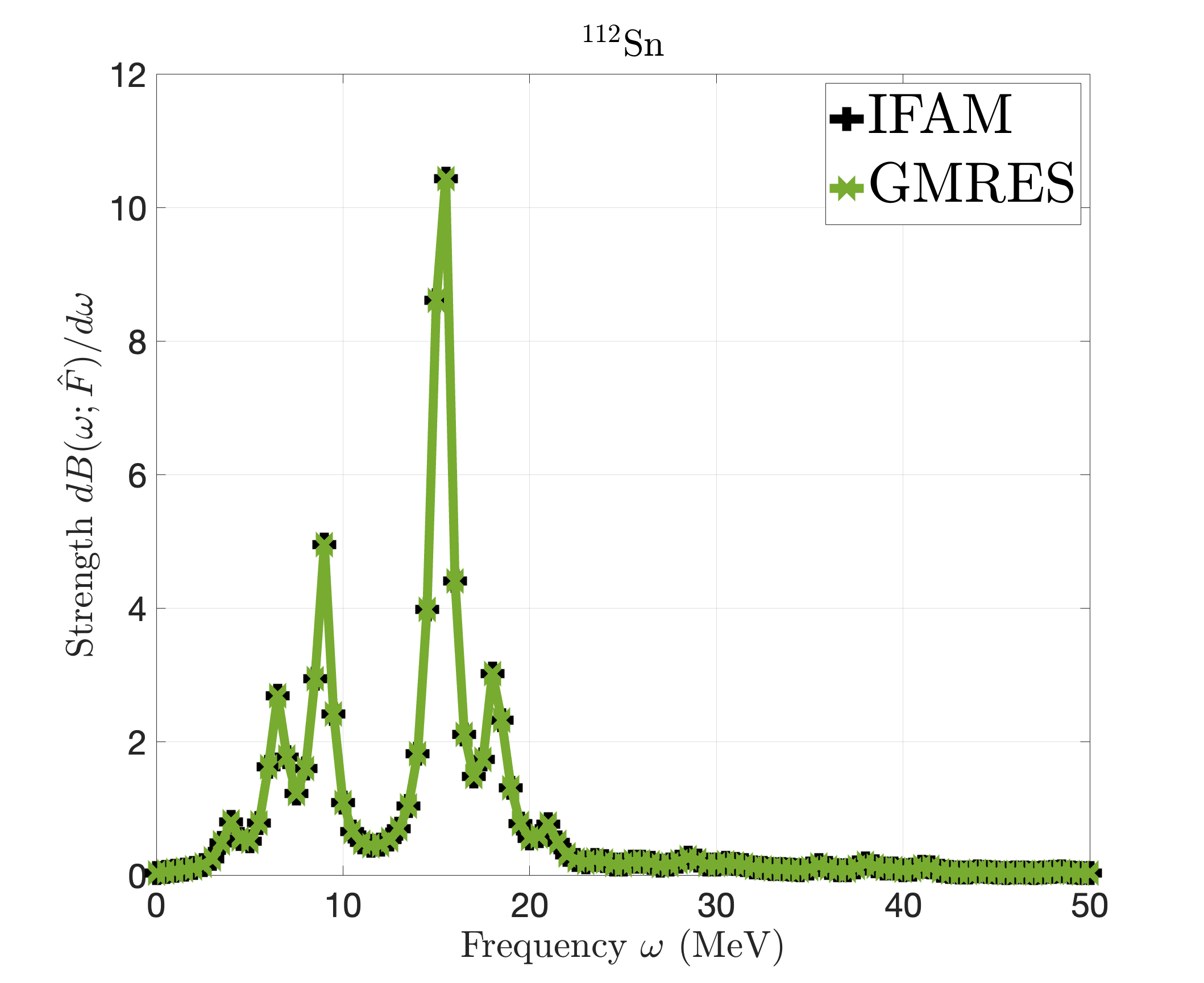}
\includegraphics[scale=0.1]{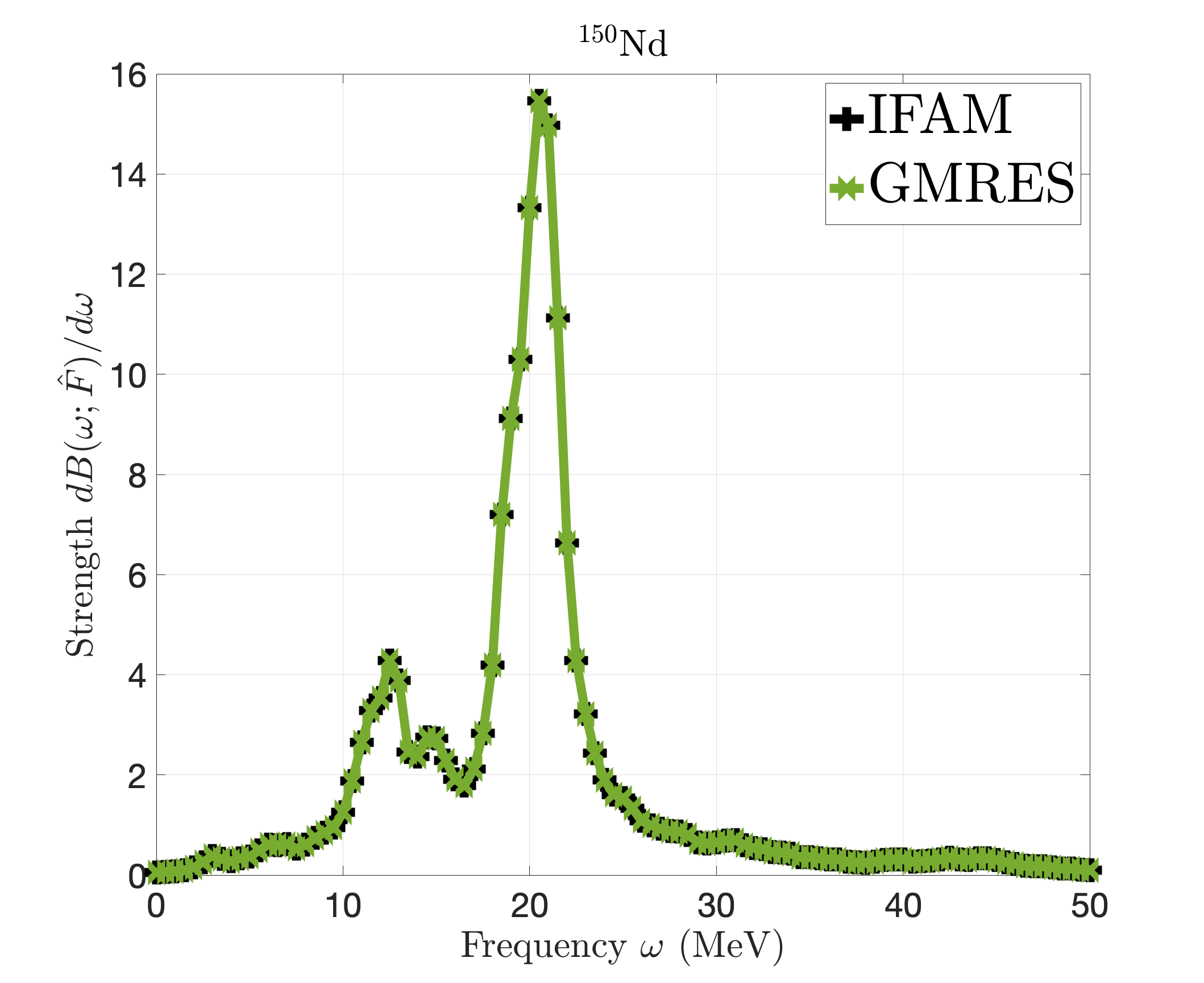}\\[-0.5em]
\includegraphics[scale=0.2]{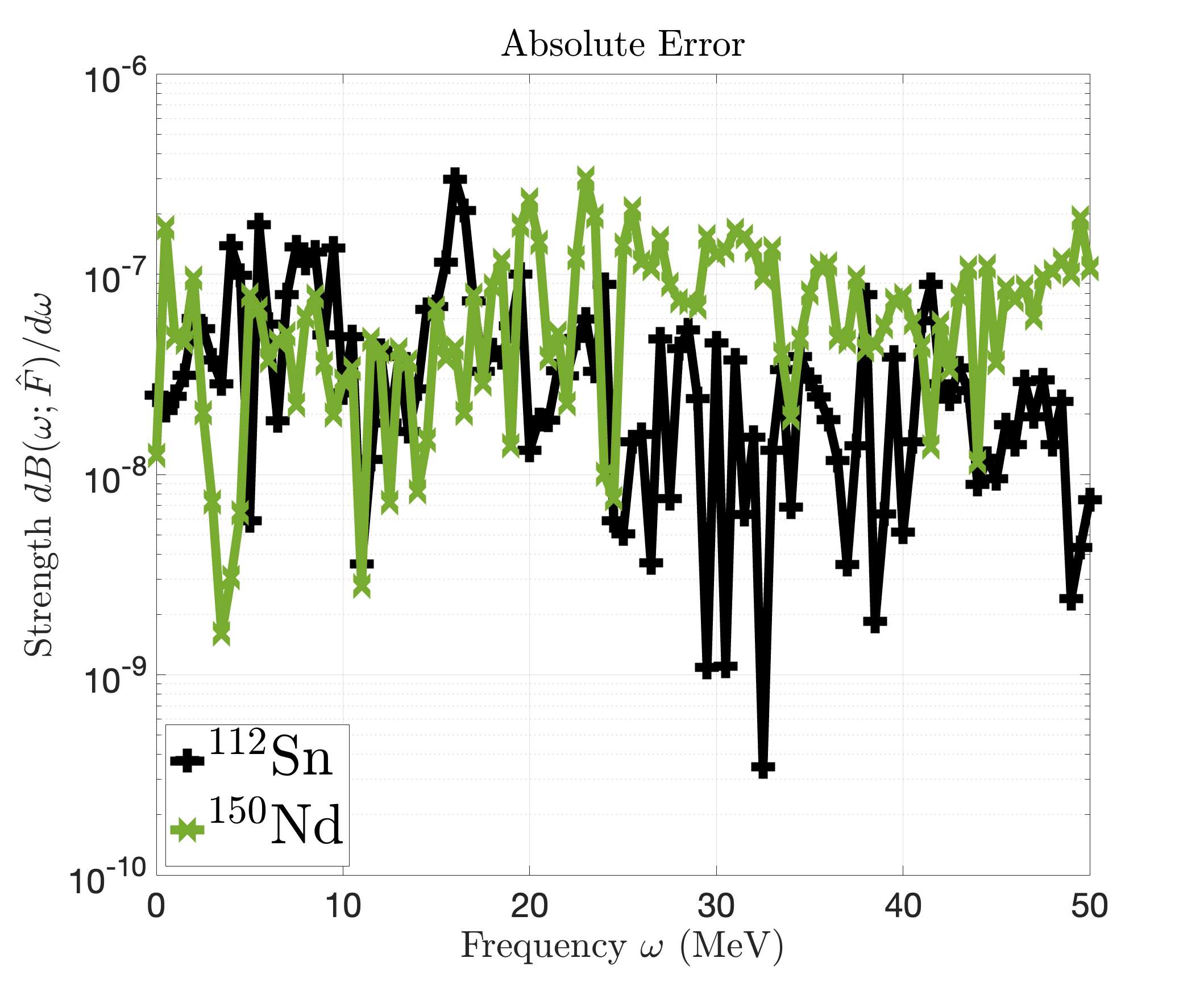}\vspace{-0.6em}
\caption{\label{fig:112Sn_150Nd_Strength_IFAM_GMRES} IFAM--GMRES comparison over $0$--$50$ MeV. Upper panels: overlaid total Gamow--Teller strengths for $^{112}$Sn (left) and $^{150}$Nd (right). Lower panel: absolute pointwise errors for $^{112}$Sn (black) and $^{150}$Nd (green).}
\end{figure}

We first compare GMRES with IFAM to identify a practical reference solution for the subsequent Lanczos comparison.
Both methods solve the same linear-response problem at each prescribed frequency, but GMRES treats it directly as a Krylov linear solve rather than as a fixed-point iteration.
Thus, when both methods are converged to the same tolerance, agreement between them is expected; the main distinction is computational efficiency.
Figure~\ref{fig:112Sn_150Nd_Strength_IFAM_GMRES} combines the comparison between GMRES and IFAM for both nuclei.
The upper panels show that the two methods produce visually indistinguishable total Gamow--Teller strength functions for $^{112}$Sn and $^{150}$Nd, while the lower panel quantifies this agreement through the absolute pointwise difference, which remains below $10^{-6}$ over the full frequency interval.

\begin{figure}[htbp]
\centering
\includegraphics[scale=0.1]{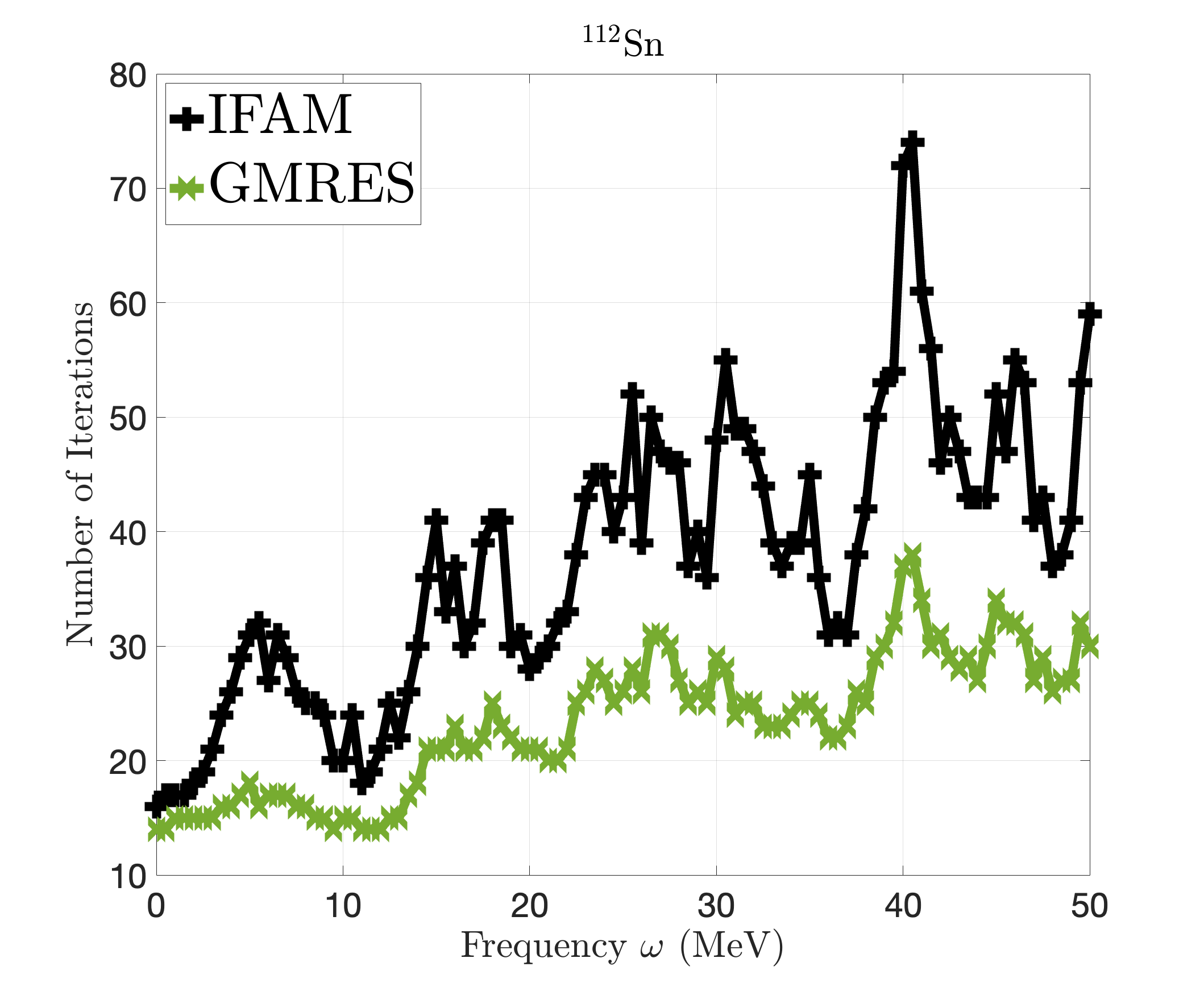}
\includegraphics[scale=0.1]{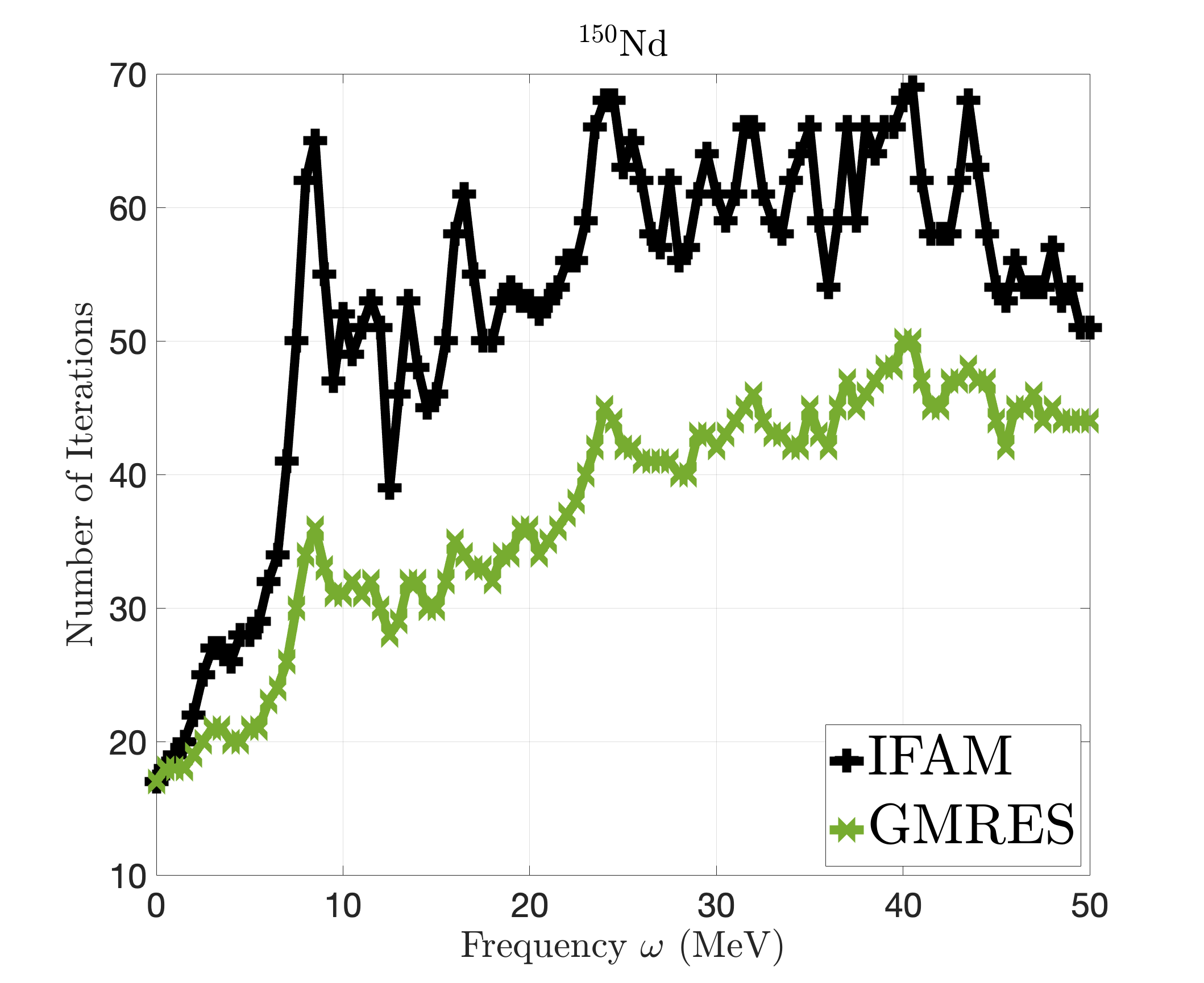}
\caption{\label{fig:time_IFAM_GMRES} Iteration counts required for IFAM and GMRES to converge for the $K=0$ component of the Gamow--Teller operator over the interval $0$--$50$ MeV. The left panel corresponds to $^{112}$Sn and the right panel to $^{150}$Nd.}
\end{figure}

The iteration-count comparison in Figure~\ref{fig:time_IFAM_GMRES} reveals a clear efficiency advantage for GMRES over the iterative FAM calculation.
This improvement is consistent with the fact that GMRES typically requires fewer Krylov iterations to reach the same convergence tolerance.
For the representative $K=0$ calculations shown in Figure~\ref{fig:time_IFAM_GMRES}, GMRES requires $2347$ total iterations for $^{112}$Sn compared with $3781$ for IFAM, and $3762$ for $^{150}$Nd compared with $5336$ for IFAM over the range $0$--$50$ MeV.

These results show that GMRES preserves the converged IFAM strength distribution while reducing the cost of the frequency-by-frequency calculation.
For this reason, in the remainder of this section we use the converged GMRES result as the frequency-by-frequency reference against which the Lanczos approximation is compared.

\subsubsection{Lanczos Approximation}

We next compare the Lanczos approximation with the GMRES reference.
Unlike GMRES and IFAM, the Lanczos method does not solve a separate linear system at each frequency.
Instead, a single Krylov projection of the reduced eigenvalue problem provides the Ritz values and approximate transition amplitudes used to reconstruct the Lorentzian-smeared strength function over the entire interval.
This subsection is organized around three questions: how the strength profile converges as the Krylov dimension increases, how close the resulting normalized distributions are to the GMRES reference, and how the runtime compares with the frequency-by-frequency methods.

\begin{figure}[htbp]
\centering
\includegraphics[scale=0.1]{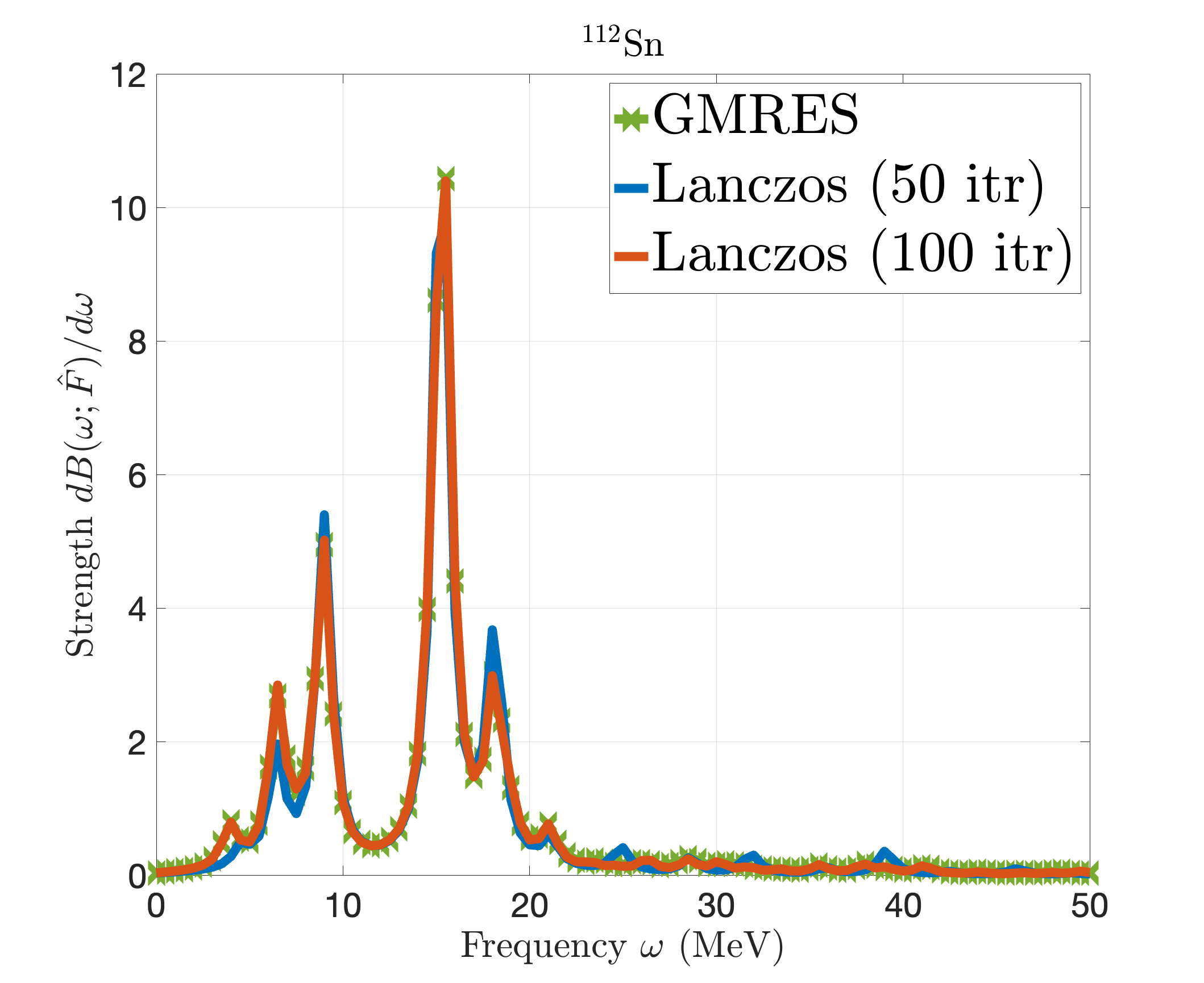}
\includegraphics[scale=0.1]{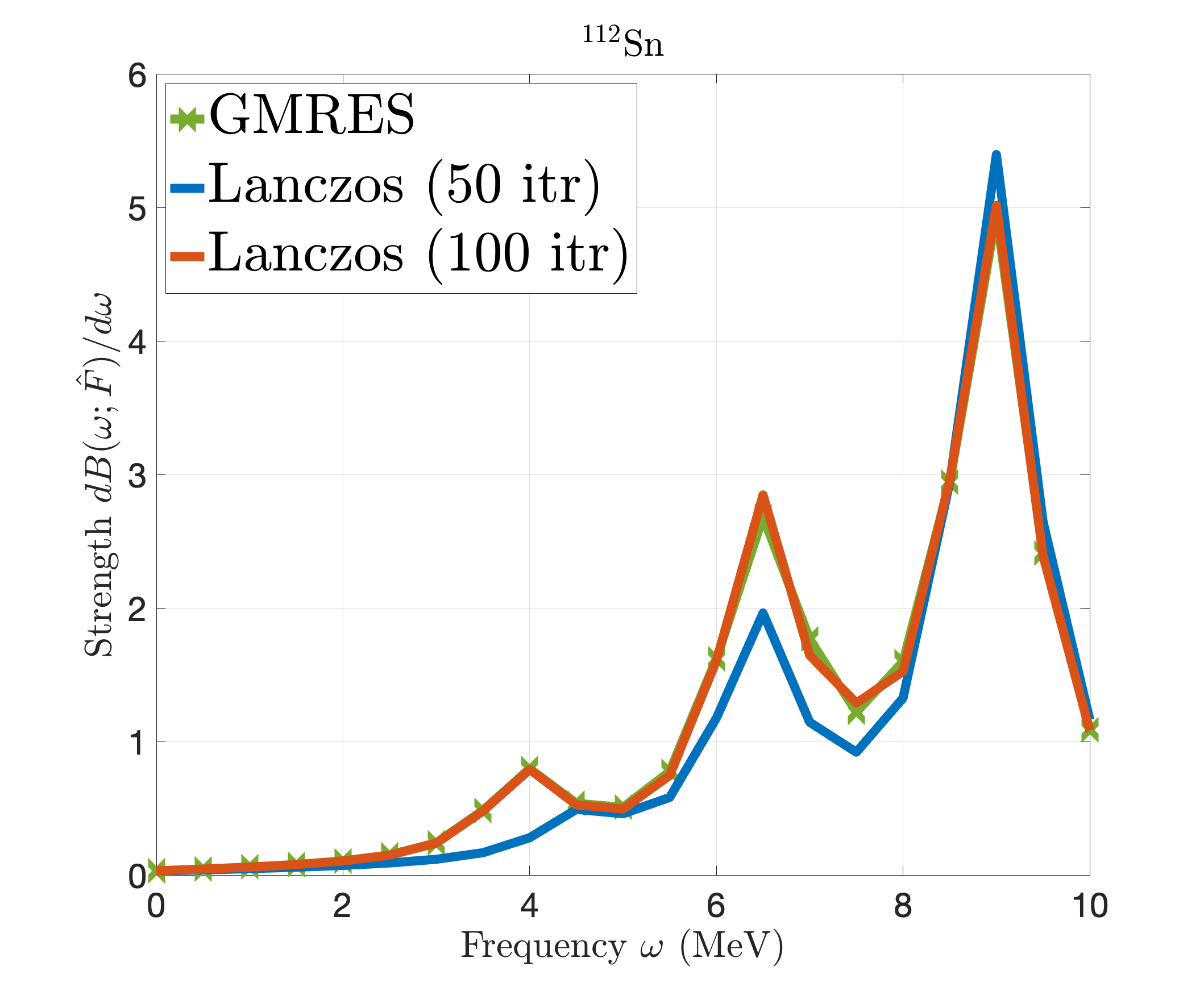}
\includegraphics[scale=0.1]{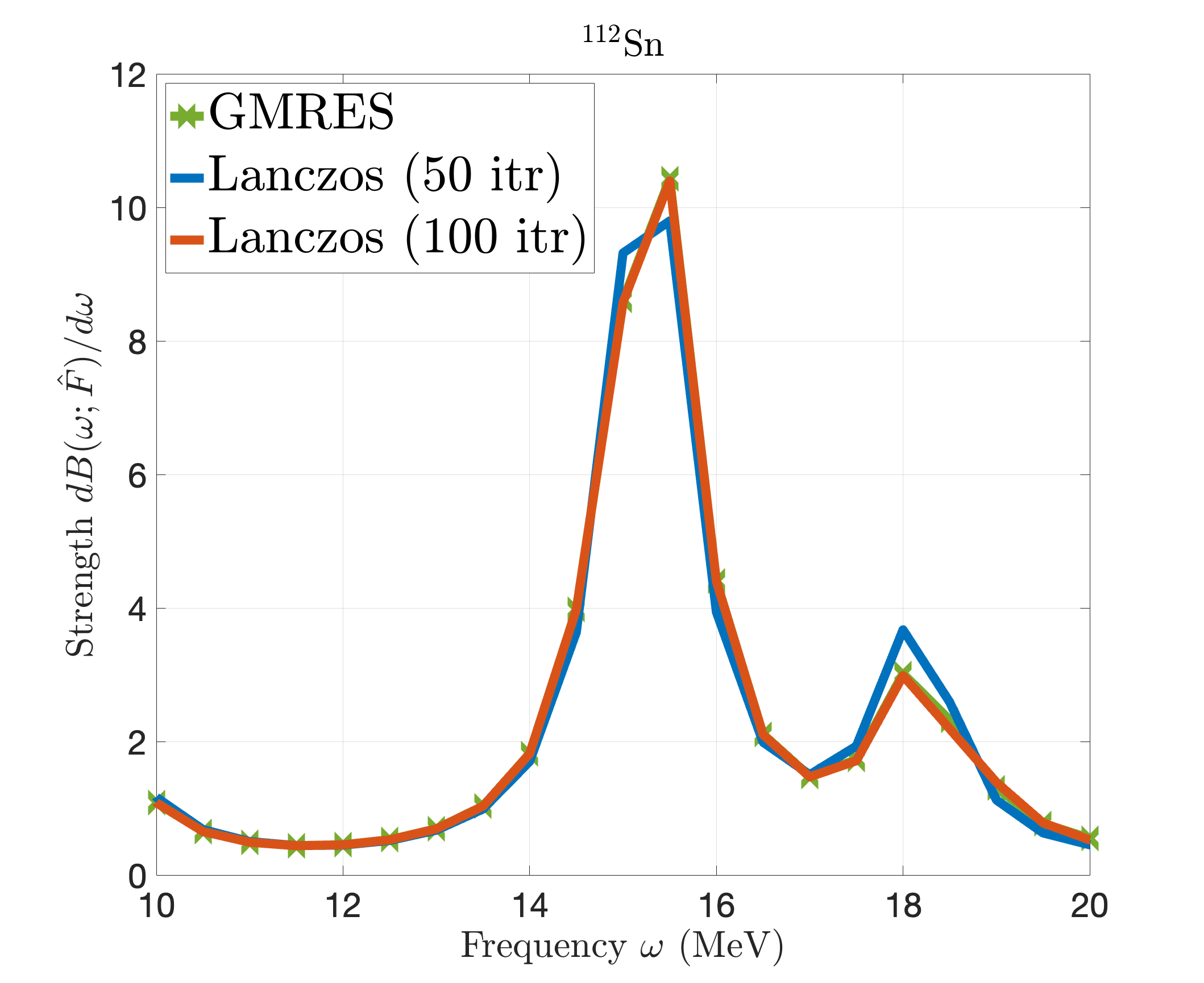}
\includegraphics[scale=0.1]{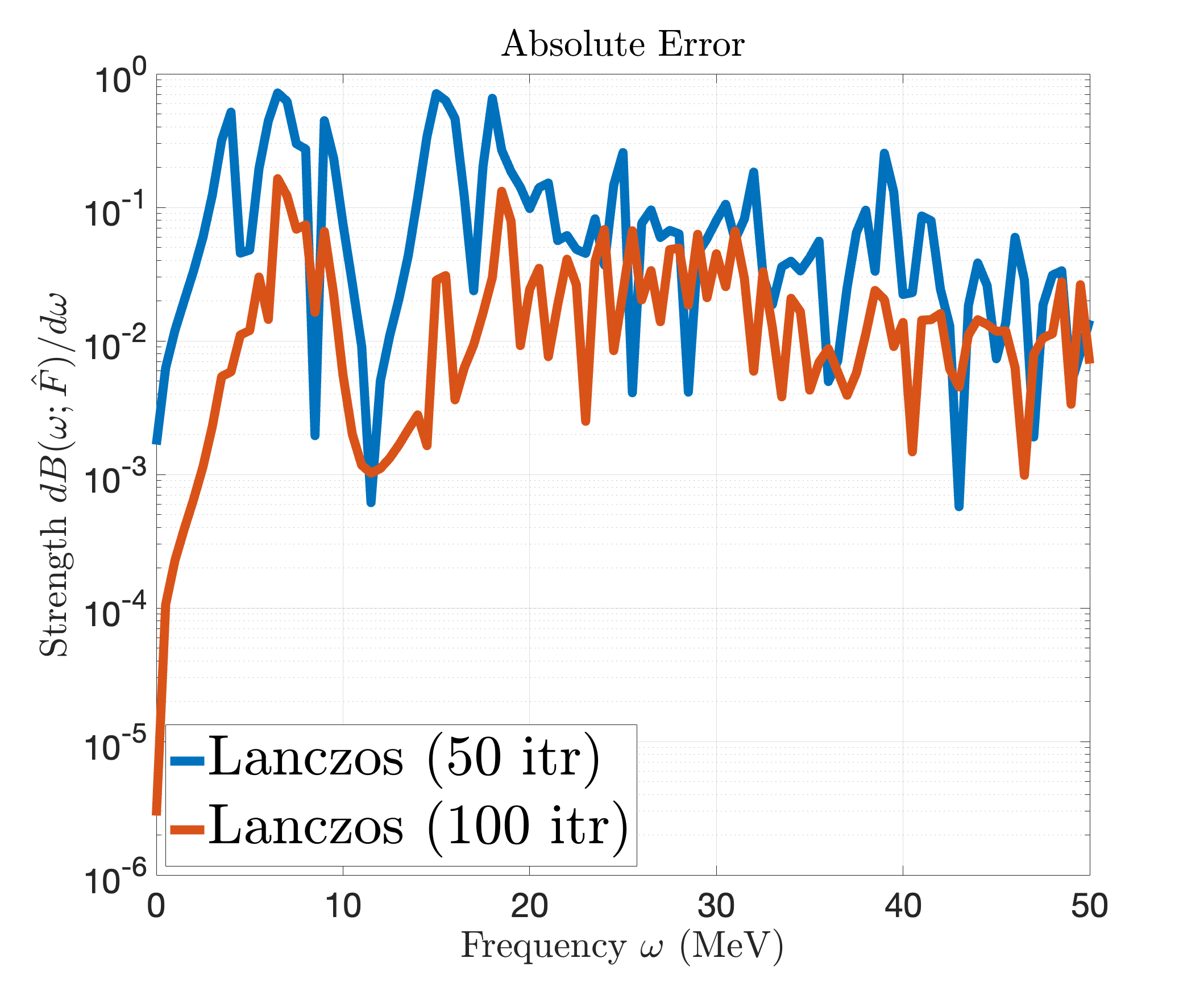}
\caption{\label{fig:112Sn_Strength} Lanczos approximation of the total Gamow--Teller strength function for $^{112}$Sn. From upper left to lower right, the panels compare the GMRES frequency-by-frequency reference with Lanczos approximations using $m=50$ and $m=100$ iterations on the full $0$--$50$ MeV interval, on the zoomed intervals $0$--$10$ MeV and $10$--$20$ MeV, and through the corresponding pointwise absolute differences from GMRES.}
\end{figure}

\paragraph{$^{112}$Sn profile convergence.}
We first consider the medium-mass nucleus $^{112}$Sn.
To illustrate convergence with respect to the Krylov dimension, we compare Lanczos approximations with $m=50$ and $m=100$ iterations against the GMRES reference.
Figure~\ref{fig:112Sn_Strength} first shows the full $0$--$50$ MeV interval and then zooms in on the lower-energy windows $0$--$10$ MeV and $10$--$20$ MeV, where the dominant peaks are located.
On the full interval, even $m=50$ captures the broad distribution of strength.
The zoomed views, however, show that $m=50$ can miss some peak heights and local structure, whereas $m=100$ gives visibly better agreement with the GMRES reference.
The pointwise absolute difference in the lower-right panel confirms this convergence trend: increasing the Lanczos dimension from $50$ to $100$ reduces the discrepancy throughout the plotted interval.

\begin{figure}[htbp]
\centering
\includegraphics[scale=0.1]{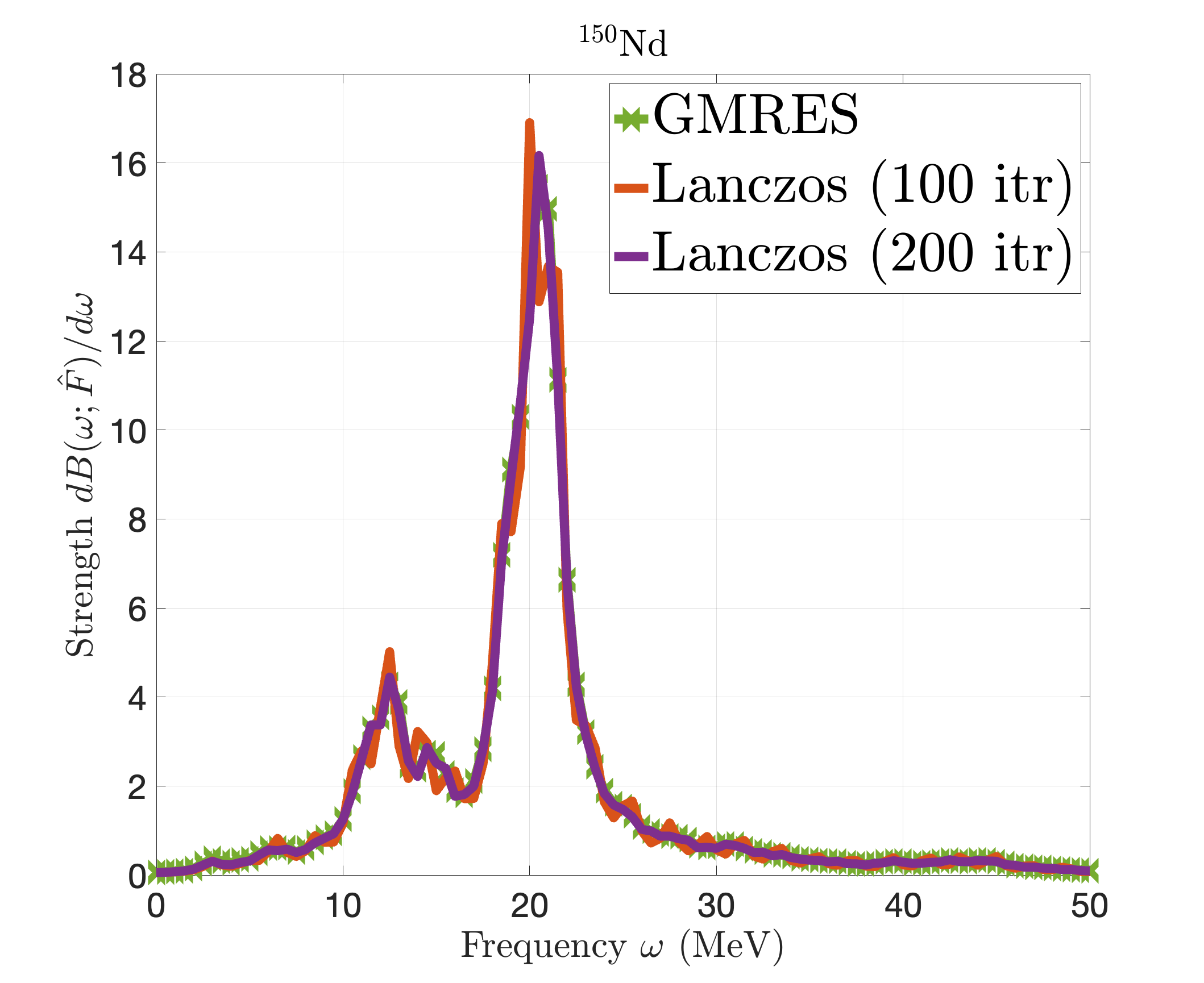}
\includegraphics[scale=0.1]{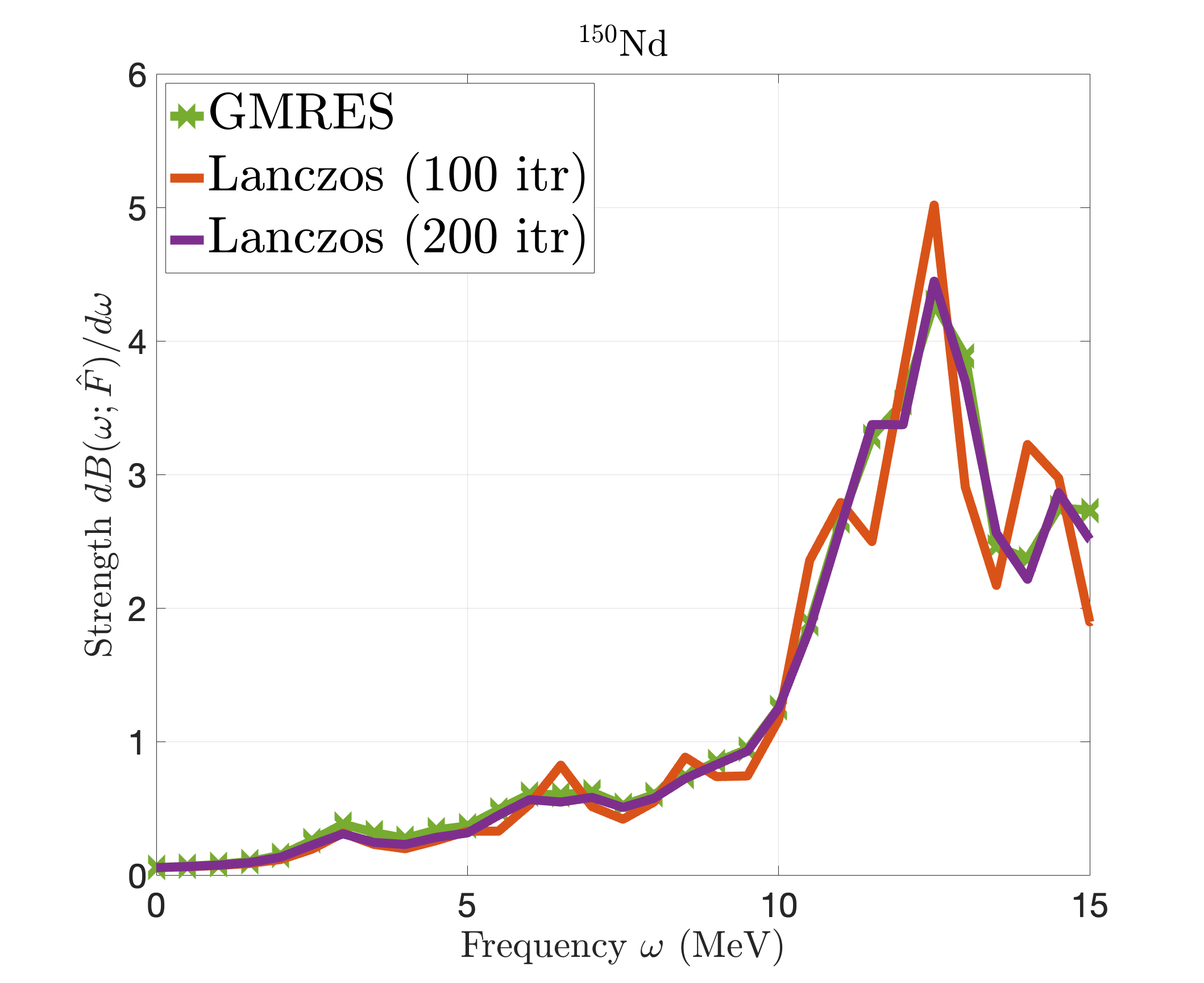}
\includegraphics[scale=0.1]{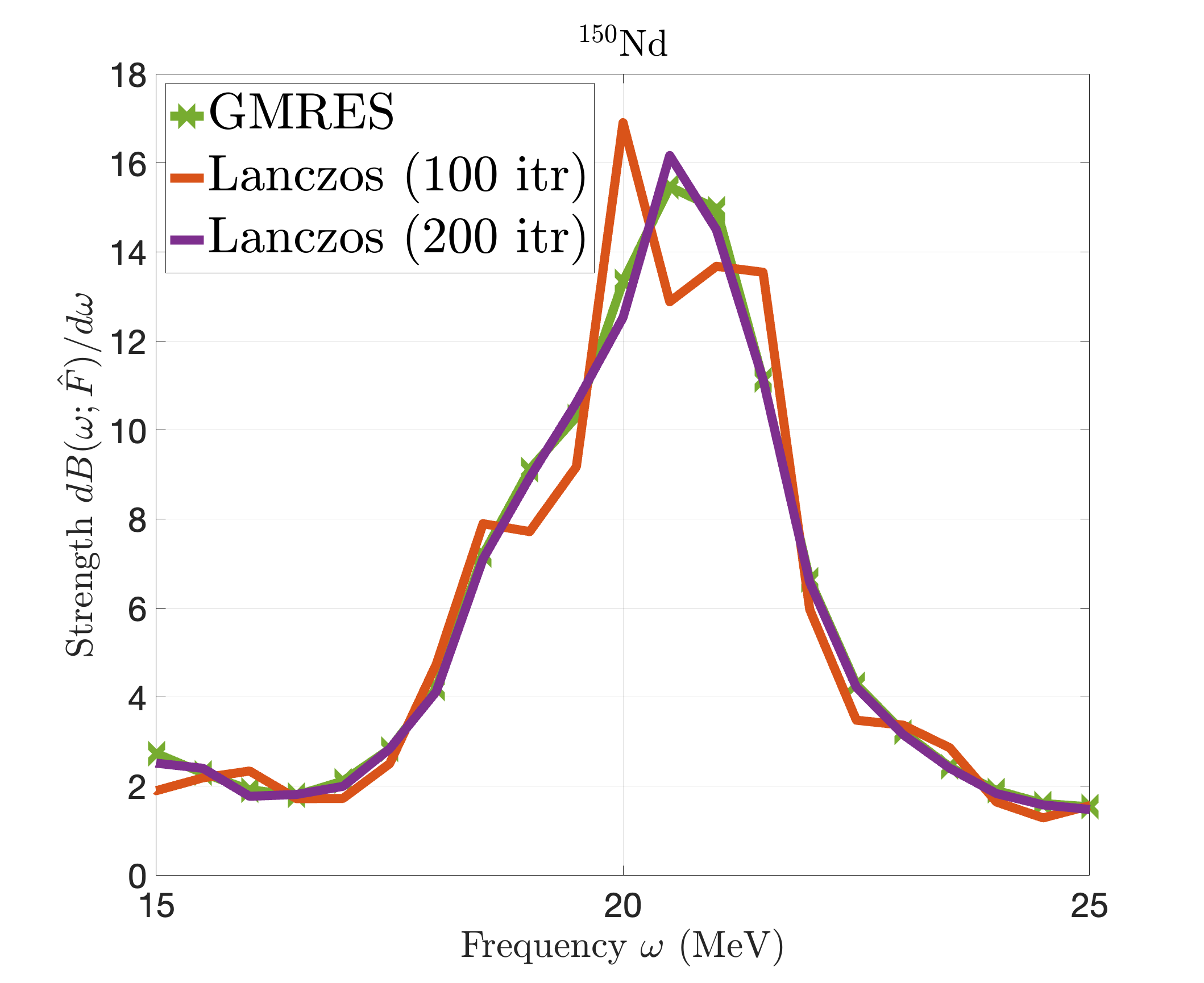}
\includegraphics[scale=0.1]{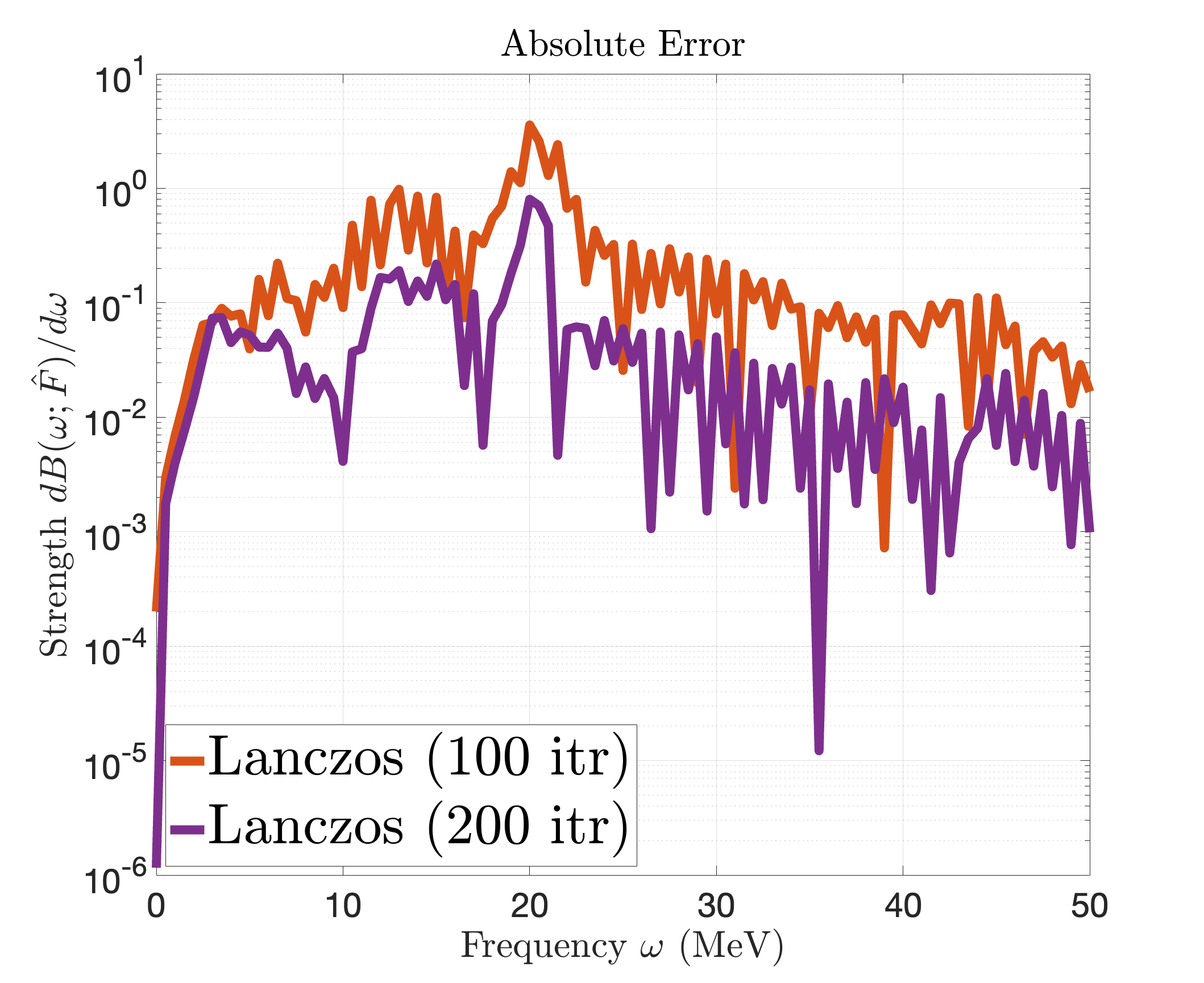}
\caption{\label{fig:150Nd_Strength} Lanczos approximation of the total Gamow--Teller strength function for $^{150}$Nd. From upper left to lower right, the panels compare the GMRES frequency-by-frequency reference with Lanczos approximations using $m=100$ and $m=200$ iterations on the full $0$--$50$ MeV interval, on the zoomed intervals $0$--$15$ MeV and $15$--$25$ MeV, and through the corresponding pointwise absolute differences from GMRES.}
\end{figure}

\paragraph{$^{150}$Nd profile convergence.}
For the heavy rare-earth nucleus $^{150}$Nd, we compare Lanczos approximations with $m=100$ and $m=200$ iterations against the GMRES reference.
Figure~\ref{fig:150Nd_Strength} first shows the full $0$--$50$ MeV interval and then zooms in on the lower-energy windows $0$--$15$ MeV and $15$--$25$ MeV, where the dominant peaks are located.
On the full interval, $m=100$ captures the broad distribution of strength, but the zoomed views show visible differences in local peak structure.
The $m=200$ reconstruction gives better agreement with GMRES in these zoomed regions.
Once again, the pointwise absolute difference in the lower-right panel confirms this convergence trend: increasing the Lanczos dimension from $100$ to $200$ reduces the discrepancy throughout the plotted interval.

\paragraph{Distribution-level error and runtime.}
As an additional distribution-level accuracy measure, we compute the Kullback--Leibler (KL) divergence between the GMRES strength distribution and each Lanczos reconstruction.
Because KL divergence is defined for probability distributions, the strength values on the sampled energy grid are first normalized separately for each nucleus as
\begin{equation}\label{eq:KL_normalization}
    p_j = \frac{S_j^{\mathrm{GMRES}}}{\sum_{\ell} S_{\ell}^{\mathrm{GMRES}}}, \qquad
    q_j^{(m)} = \frac{S_j^{\mathrm{Lanczos},m}}{\sum_{\ell} S_{\ell}^{\mathrm{Lanczos},m}}.
\end{equation}
We then evaluate
\begin{equation}\label{eq:KL_divergence}
    D_{\mathrm{KL}}\left(p\middle\|q^{(m)}\right)
    = \sum_j p_j \log\left(\frac{p_j}{q_j^{(m)}}\right),
\end{equation}
where the GMRES distribution $p$ is treated as the reference, or ``true,'' distribution for the corresponding nucleus.
The KL divergence is nonnegative and equals zero only when the two normalized distributions agree exactly; smaller values therefore indicate better agreement with the GMRES reference.
The values in Table~\ref{tab:KL_divergence} provide the same type of convergence evidence for both $^{112}$Sn and $^{150}$Nd: increasing the Lanczos dimension substantially reduces the distribution-level discrepancy.
For $^{112}$Sn, the KL divergence decreases by more than an order of magnitude from $m=50$ to $m=100$.
For $^{150}$Nd, the KL value at $m=200$ is likewise much smaller than the value at $m=100$, consistent with improved convergence as the Krylov dimension is increased.

\begin{table}[htbp]
\centering
\begin{tabular}{ccc}
\hline
Nucleus & Lanczos iterations & $D_{\mathrm{KL}}(p_{\mathrm{GMRES}}\|q_{\mathrm{Lanczos}})$ \\
\hline
$^{112}$Sn & $m=50$ & $2.8958\times 10^{-2}$ \\
$^{112}$Sn & $m=100$ & $2.3455\times 10^{-3}$ \\
\hline
$^{150}$Nd & $m=100$ & $1.6739\times 10^{-2}$ \\
$^{150}$Nd & $m=200$ & $8.4926\times 10^{-4}$ \\
\hline
\end{tabular}
\caption{KL divergence between the normalized GMRES strength distribution and normalized Lanczos reconstructions for $^{112}$Sn and $^{150}$Nd. Smaller values indicate closer agreement with the GMRES reference.}\label{tab:KL_divergence}
\end{table}

\begin{figure}[htbp]
\centering
\includegraphics[scale=0.2]{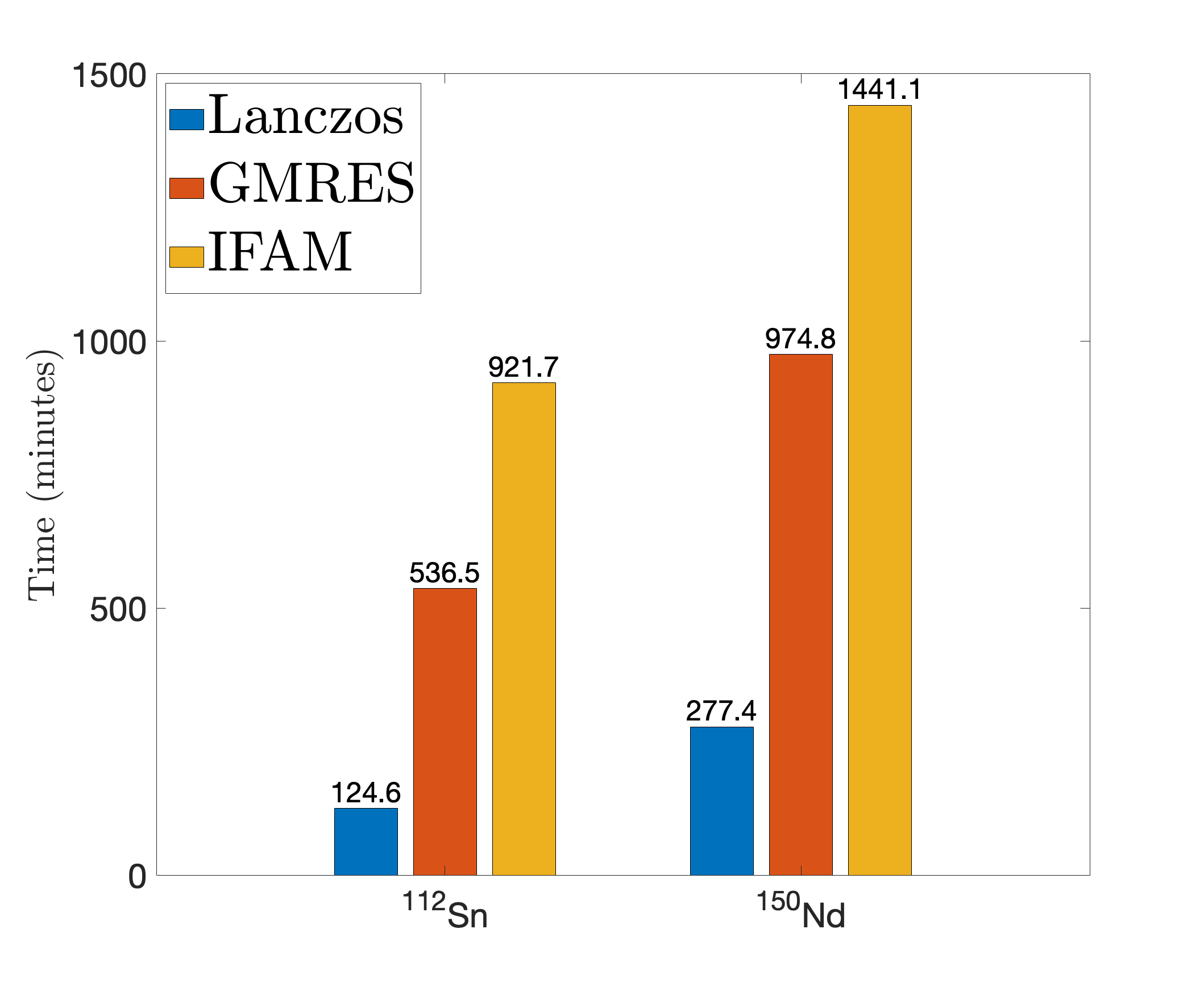}
\caption{\label{fig:time} Elapsed runtime, in minutes, of IFAM, GMRES, and Lanczos for the Gamow--Teller $0$--$50$ MeV strength-function calculations in $^{112}$Sn and $^{150}$Nd. The Lanczos timings use $m=100$ for $^{112}$Sn and $m=200$ for $^{150}$Nd.}
\end{figure}

Together, these comparisons isolate the convergence behavior of the Lanczos reconstruction for the two realistic test nuclei.
The runtime comparison in Figure~\ref{fig:time} shows the corresponding efficiency differences among IFAM, GMRES, and Lanczos.
The improved runtime of GMRES relative to IFAM is consistent with its smaller iteration count: for the representative $K=0$ calculations in Figure~\ref{fig:time_IFAM_GMRES}, GMRES requires $2347$ total iterations for $^{112}$Sn compared with $3781$ for IFAM, and $3762$ for $^{150}$Nd compared with $5336$ for IFAM over the range $0$--$50$ MeV.
The Lanczos timings can be interpreted through the matvec-equivalent cost model in Section~\ref{subsec:matvecs}.
Including the additional products needed to form the transition amplitudes, an $m$-step Lanczos calculation requires $6m$ matvecs.
Thus, the Lanczos runs used for the timing comparison cost $600$ matvecs for $^{112}$Sn with $m=100$ and $1200$ matvecs for $^{150}$Nd with $m=200$.
In the same bookkeeping, IFAM and GMRES require one matvec-equivalent operation per iteration, so the corresponding GMRES costs are $2347$ and $3762$ matvecs, respectively.
These estimates explain why Lanczos remains faster than the frequency-by-frequency solvers in Figure~\ref{fig:time}, even though the precise wall-clock speedups also depend on implementation overhead and post-processing costs.

\section{Conclusion}

We have presented a symmetric Lanczos framework for approximating charge-changing QRPA strength functions in a matrix-free setting. Starting from the QRPA linear-response equation and passing through eigendecomposition-based and reduced-eigenproblem formulations, we obtained a representation of the strength function that is well suited to Krylov projection. The resulting method differs conceptually from conventional frequency-by-frequency FAM calculations in that a single Lanczos run captures spectral information across an entire energy interval.

The numerical experiments show that GMRES provides an efficient frequency-by-frequency reference for the medium-mass and heavy nuclei considered here, reproducing converged IFAM strength profiles while requiring fewer iterations. Against this reference, the Lanczos approximation reproduces the same overall Gamow--Teller strength profiles for $^{112}$Sn and $^{150}$Nd while requiring substantially fewer matvec-equivalent operations.

Taken together, these results indicate that the symmetric Lanczos method provides a practical alternative for QRPA strength-function calculations when broad spectral information is needed. The approach is especially attractive for deformed and heavy nuclei, where explicit QRPA matrices are too large to build and repeated frequency-by-frequency solves become expensive. Natural directions for future work include extending the method to additional external fields, refining the implementation of the reduced problem, and exploring larger systematic calculations in realistic nuclear-structure applications.

\bibliographystyle{plain}
\bibliography{refs}

\end{document}